\documentclass[a4paper,11pt]{article}
\usepackage{jheppub} 
\usepackage{lineno}
\usepackage{subcaption}
\usepackage{graphicx}

\newcommand{\be}{\begin{equation}}
\newcommand{\ee}{\end{equation}}

\newcommand{\bs}{\begin{equation}\begin{aligned}}
\newcommand{\es}{\end{aligned}\end{equation}}

\newcommand{\bg}{\begin{gathered}}
\newcommand{\eg}{\end{gathered}}

\title{\boldmath A complex scalar field theory for charged fluids, superfluids, and fracton fluids }

\author{Aleksander G\l{}\'{o}dkowski}
\affiliation{Institute of Theoretical Physics, Wroc\l{}aw  University  of  Science  and  Technology,  50-370  Wroc\l{}aw,  Poland}

\emailAdd{aleksander.glodkowski@pwr.edu.pl}

\abstract{We propose a field-theoretic framework for ideal hydrodynamics of charged relativistic fluids formulated in terms of a complex scalar field defined on a spacelike hypersurface comoving with the fluid. In the normal phase, the dynamics of charge-carrying fluids is constrained by the restrictive chemical shift symmetry, which locks charges to fixed positions in the comoving plane as they are transported through space by the fluid's motion. On the other hand, in the superfluid phase, the chemical shift symmetry is relaxed to a constant shift, allowing charges to redistribute freely across the comoving hypersurface. We demonstrate that both models recover the respective nonlinear hydrodynamic equations and provide explicit expressions for the collective variables of hydrodynamics in terms of the theory’s fields. Introduced models provide a UV completion to the effective field theories of hydrodynamics constructed in terms of the Goldstone fields. Finally, we propose a relativistic fracton fluid phase as a natural interpolation between the normal and superfluid phases, in which the mobility of elementary charges is constrained by a linear shift symmetry in the comoving space.}

\begin{document}
\maketitle
\flushbottom

\section{Introduction}

Fluid dynamics is one of the oldest and most established fields of physics, with applicability ranging from microscopic quantum fluids, such as the quark-gluon plasma, all the way to the universe itself. Despite its long and illustrious history, a complete understanding of fluids is still lacking. Some of the most pressing open problems include nonlinear phenomena such as turbulence and the million-dollar problem on the Navier–Stokes equations, but also the behavior of strongly correlated quantum fluids, and questions concerning their existence (or lack thereof) at zero temperature \cite{Endlich:2010hf,Dersy:2022kjd,Cuomo:2024ekf}. 

As of today, hydrodynamics is generically understood as the most general set of dynamical equations consistent with the symmetries of a system and the laws of thermodynamics, expressed in terms of collective hydrodynamic variables. In writing these equations, one remains somewhat agnostic about the finer details of the system, such as the interactions of its microscopic constituents, which are encoded in a finite set of phenomenological transport coefficients. This paradigm has proven extremely successful owing to its robustness, universality, and broad applicability in describing the low-energy dynamics of many-body systems. Nevertheless, the utility of this approach remains somewhat limited in scope, particularly when it comes to addressing some of the open problems mentioned above.

From a theoretical standpoint, a more compelling picture is offered by a field-theoretic formulation of hydrodynamics in terms of a local action \cite{Son:2002zn,Dubovsky:2005xd,Dubovsky:2011sj,Nicolis:2011cs,Kovtun:2014hpa,Crossley:2015evo}. In this framework, the hydrodynamic equations follow from the Euler–Lagrange equations for the relevant low-energy fields, while the path-integral formalism grants access to the correlation functions and scattering amplitudes (see, e.g., \cite{PhysRevLett.94.175301,Endlich:2010hf,Nicolis:2011cs,Dersy:2022kjd,Jain:2023obu,Delacretaz:2023ypv,Cuomo:2024ekf}). In short, the action formulation allows the use of the full quantum field theory (QFT) toolkit. The models in question are effective field theories (EFTs) where infrared degrees of freedom are the Nambu-Goldstone bosons of spontaneously broken symmetries. Unfortunately, without microscopic input from the UV, these EFTs can suffer from various pathologies.

To illustrate this, let us consider a $U(1)$ superfluid at zero temperature. Its long-wavelength dynamics is governed by a single scalar Goldstone field $\psi$ transforming nonlinearly under the action of the broken $U(1)$ symmetry, $\psi \rightarrow \psi + \lambda$. The corresponding effective action takes the following general form 
\be \label{eq:superfluidEFT}
S = \int d^{d+1} x \sqrt{-g}  \, \mathcal L(X)\,, \quad X \equiv D_\mu \psi D^\mu \psi\,,
\ee 
 where $D_\mu \psi = \partial_\mu \psi - A_\mu$ is a covariant derivative and $\mathcal L(X)$ is an arbitrary function. For example, choosing $\mathcal L(X) = X$ gives a simple quadratic theory with linear equations of motion. In general, however, the resulting equations are nonlinear and lead to the formation of caustic singularities in finite time, wherein the characteristics intersect and second-order derivatives diverge, signaling the breakdown of the EFT \cite{Babichev:2016hys}. These issues can be circumvented by introducing a UV completion in terms of a complex scalar field, which reproduces the same low-energy dynamics as the EFT model (up to the formation of caustic singularities in the EFT) while regularizing the caustic instability \cite{Babichev:2017lrx}.

For superfluids, a suitable UV completion is given by a canonical complex scalar field action \cite{Babichev:2018twg}
\be \label{eq:superfluidMicroscopic}
S = \int d^{d+1} x \sqrt{-g} \, \Big( \frac{1}{2} |D_\mu \Phi|^2 - V(|\Phi|) \Big)\,, \quad  V(|\Phi|) = - \frac{m^2}{2} |\Phi|^2 - \frac{\lambda}{4} |\Phi|^4\,,
\ee
where $D_\mu \Phi = \left( \partial_\mu - i A_\mu \right) \Phi$. By placing the system at finite chemical potential via a constant background gauge field $A_\mu = (\mu_0,0)$ with $\mu_0 > m$, and implementing the polar decomposition $\Phi = \sqrt{\rho} e^\psi$, the theory Eq.~\eqref{eq:superfluidMicroscopic} reproduces Eq.~\eqref{eq:superfluidEFT} in the long-wavelength limit after integrating out the Higgs mode \cite{Joyce:2022ydd}.
Clearly, the theory Eq.~\eqref{eq:superfluidMicroscopic} contains more information than the EFT Eq.~\eqref{eq:superfluidEFT}. For example, it can be used to compute the healing length of the superfluid. It is precisely the absence of information about the short-range physics in the EFT Eq.~\eqref{eq:superfluidEFT} that manifests itself in the appearance of caustic singularities.

While the UV completion of the superfluid EFT at zero temperature is well understood, the situation becomes more subtle at finite temperatures, where thermal excitations populate the normal component and give rise to a two-fluid description consisting of a (quantum) superfluid coupled to a normal (classical) fluid. An analogous question can be posed regarding the UV completion of charged fluids in the normal phase, wherein the charge and thermal sectors are likewise intertwined. 

Motivated by these considerations, in this manuscript, we propose a UV completion to the EFTs of charged fluids \cite{Dubovsky:2011sj} and superfluids at finite-temperature \cite{Nicolis:2011cs}. For this purpose, we utilize a geometric picture of fluid flow in terms of a comoving spatial hypersurface and define on it a complex scalar field transforming in the linear representation of the $U(1)_Q$ charge symmetry. The models proposed herein reproduce the established constitutive relations for the ideal part of hydrodynamic currents, and the hydrodynamic excitations are identified with the Nambu-Goldstone modes corresponding to small fluctuations around the state of thermodynamic equilibrium. 

We show that ordinary charged fluids are necessarily invariant under a restrictive \textit{chemical shift} symmetry that freezes the dynamics of the elementary charges in the comoving space. In this interpretation, charges exhibit fractonic immobility on the comoving plane but are dragged through spacetime by the fluid flow. This symmetry is relaxed in the superfluid phase in order to account for the dynamics of the superfluid component. Our formulation naturally accommodates a classification of exotic fluid phases interpolating between normal fluids and superfluids. These \textit{fracton fluids} exhibit partially restricted mobility, a consequence of the comoving multipole symmetries\footnote{Analogous ''crystal-multipole'' symmetries were introduced in \cite{Jain:2024ngx}.}, which in our approach emerge as a finite truncation of the chemical shift symmetries.

The plan for the paper is as follows. In Sec.~\ref{sec:comoving} we review the comoving formulation of fluid flow and discuss the geometry of the comoving hypersurface. Then, in Sec.~\ref{sec:charged} we show that perfect charged fluids exhibit the restrictive chemical shift symmetry and propose a complex scalar field theory describing their hydrodynamics. Next, in Sec.~\ref{sec:superfluid} we study the superfluid phase, in which the chemical shift symmetry is supplanted by by the standard $U(1)_Q$ shifts and formulate a theory for finite-temperature superfluids composed of normal and superfluid components. Sec.~\ref{sec:fracton} is devoted to the study of fracton fluids, which are introduced as fluid phases interpolating between the normal and superfluid states. We conclude with a discussion of our results in Sec.~\ref{sec:conclusion} where we place our findings in a broader perspective and outline possible extensions of the framework. 



\textbf{Notation and conventions.} For concreteness of presentation, throughout this paper we specialize to $2+1$--spacetime dimensions, although most (if not all) of the results can be straightforwardly generalized to $3+1$--dimensions. We adopt the mostly-plus convention for the Minkowski metric, $\eta_{\mu \nu} = \text{diag}(-,+,+)$, use natural units and employ Einstein summation convention. 
We also use round brackets to denote total symmetrization in the indices
\be 
A^{(I_1 \dots I_N)} =\frac{1}{N!} \sum_\pi A^{I_{\pi(1)} \dots I_{\pi(N)}} \,,
\ee 
whereas vertical bars $| \cdot |$ exclude the indices from the symmetrization. For example 
\be 
A^{(I_1 |I_2| \dots I_N)} =\frac{1}{(N-1)!} \sum_\pi A^{I_{\pi(1)} I_2 \dots I_{\pi(N)}} \,,
\ee 
where the sum is taken over all permutations of the indices $\{ I_1, I_3, \dots I_N \}$, leaving $I_2$ fixed. 

\section{Comoving formulation of fluid dynamics}\label{sec:comoving}
In this section we review the comoving formulation of perfect fluids and encode their relativistic hydrodynamics in an action principle.
The variational approach to fluid mechanics in terms of a set of comoving coordinates has a long history~\cite{10.1063/1.1706053,annurev:/content/journals/10.1146/annurev.fl.20.010188.001301,Bahcall:1991an}, see also~\cite{jackiw2002lectures} for a pedagogical introduction. More recently, the problem has been revisited from a modern effective field theory viewpoint~\cite{Dubovsky:2005xd}, which we will mostly follow in this section. We also provide a complementary geometrical interpretation of the comoving formulation in terms of spacelike hypersurface embedded in the physical spacetime. Our setup aligns closely with the geometric framework of elasticity theory introduced in~\cite{Armas:2019sbe}. Although the comoving hypersurface perspective may currently appear as a mere reinterpretation, it sets the stage for the constructions developed in the upcoming sections.

\subsection{Reshuffling the fluid}

A classical state of an ensemble of $N$ point particles is specified by the positions and velocities of the particles, each labelled by a discrete index $k=1, \dots, N$. In a continuous medium $N \rightarrow \infty$, these labels are promoted to internal coordinates $\phi^{I=1,2}$, such that the finite set of particle positions $\mathbf X_k(t)$ is replaced by a continuous function $\mathbf X(\phi^I, t)$. For fixed $\phi^I$, the function $\mathbf X(\phi^I, t)$ traces out the worldline of the fluid parcel labelled by the coordinates $\phi^I$. Of course, there is a huge amount of gauge freedom in choosing the $\phi^I$ coordinates, corresponding to the relabelling of the fluid elements, which we emphasize is not a physical operation. This freedom can be fixed by choosing $\phi^I$ to coincide with the positions of the fluid elements at some reference instance of time. 

It is then possible to invert $\mathbf X(\phi^I, t)$ to obtain a set of scalar fields $\phi^I(x,t)$ defined over spacetime. These scalar fields constitute the dynamical coarse-grained degrees of freedom that parameterize the continuum medium and which we will employ to construct the effective field theory. In order to describe a fluid phase\footnote{For a solid, one only demands invariance under translations and rotations of the comoving coordinates.}, we demand that the dynamics be invariant under area-preserving diffeomorphisms (APDs) of the comoving coordinates 
\be \label{eq:diffeo}
\phi^I \rightarrow  \tilde \phi^I(\phi)\,, \quad \Big| \frac{\partial \tilde \phi}{\partial \phi}\Big| = 1\,,
\ee
which we will denote as $\text{SDiff}(\mathbb R^2)$. 
Importantly, the global symmetry Eq.~\eqref{eq:diffeo} reflects the invariance of fluid dynamics under physical reshuffling of fluid elements, in contrast to the mere relabelling gauge freedom discussed earlier. This invariance stems from the absence of shear forces in an ideal fluid. Meanwhile, the condition of area preservation encodes the fluid's resistance to compression and dilation, which changes the physical state of the fluid (and costs energy). Without loss of generality, any infinitesimal APD Eq. \eqref{eq:diffeo} can be written as
\be \label{eq:APDs}
\phi^I \rightarrow \phi^I + \epsilon^{IJ}\frac{\partial \Sigma (\phi) }{\partial \phi^J}\,,
\ee 
for some arbitrary function $\Sigma (\phi)$. To get some control over the unwieldy set of APDs, it is useful to decompose $\Sigma(\phi)$ into a polynomial basis
\be \label{eq:diffPolynomial}
\Sigma (\phi) = \Sigma_0 + \Sigma_I \phi^I + \frac{1}{2} \Sigma_{IJ} \phi^I \phi^J + \cdots \,
\ee  
where the component of degree $N$ is denoted by
\be
\Sigma_N(\phi) = \frac{1}{N} \Sigma_{I_1 \dots I_N} \phi^{I_1} \dots \phi^{I_N}.
\ee  
Each such component generates an independent infinitesimal transformation
\be \label{eq:SigmaN}
\delta_{\Sigma_N} \phi^I = \epsilon^{IJ} \frac{\partial \Sigma_N}{\partial \phi^J} = \epsilon^{I I_1} \Sigma_{I_1 \dots I_N} \phi^{I_2} \dots \phi^{I_N} \,,
\ee 
parametrized with the symmetric tensor $\Sigma_{I_1 \dots I_N}$. 
These transformations form an infinite-dimensional Lie algebra
\be 
[\delta_{\Sigma_N}, \delta_{\Sigma_{N^\prime}}] = \delta_{[\Sigma_N, \Sigma_{N^\prime}]} \,, 
\ee 
with the Lie bracket defined as 
\be 
[\Sigma_N, \Sigma_{N^\prime}] = \epsilon^{IJ} \frac{\partial \Sigma_{N^\prime}}{\partial \phi^I} \frac{\partial \Sigma_N  }{\partial \phi^J}\,.
\ee 
Expanding this, one finds
\be
[\Sigma_N, \Sigma_{N^\prime}] =  \epsilon^{IJ} \Sigma^\prime_{ I I_1 \dots I_{N^\prime-1} } \Sigma_{J I_{N^\prime}  \dots I_{N^{\prime \prime}}} \phi^{I_1} \dots \phi^{I_{N^{\prime \prime}}}\,,
\ee 
where $N^{\prime \prime} = N + N^\prime - 2$. In other words,
\be 
[\delta_{\Sigma_N}, \delta_{\Sigma_{N^\prime}}] = \delta_{\Sigma_{N^{\prime \prime}}} \,, 
\ee 
with parameter 
\be 
\Sigma^{\prime \prime}_{I_1 \dots I_{N^{\prime \prime}}} = N^{\prime \prime} \epsilon^{I J}  \Sigma^\prime_{ I I_1 \dots I_{N^\prime-1} } \Sigma_{J I_{N^\prime}  \dots I_{N^{\prime \prime}}}\,.
\ee 
Let us introduce a basis of generators spanning the Lie algebra
\be \label{eq:generatorsF}
F^{I_1 \dots I_N} \equiv \frac{1}{N} \epsilon^{JK} \partial_K P^{I_1 \dots I_N} \partial_J = \epsilon^{J (I_1} \phi^{I_2} \dots \phi^{I_N )} \partial_J\,,
\ee
where $P^{I_1 \dots I_N} = \phi^{I_1} \dots \phi^{I_N}$ are the homogenous polynomials of degree $N$ and $\partial_I = \frac{\partial}{\partial \phi^I}$. 
In this basis, the elements of the Lie algebra are expressed as 
\be 
\delta_{\Sigma_N} = \Sigma_{I_1 \dots I_N}  F^{I_1 \dots I_N} \,,
\ee 
and the structure of the algebra is encoded in the commutation relations
\bs 
[F^{I_1 \dots I_N}, F^{J_1 \dots J_{N^\prime}}] &= N^{\prime \prime} \epsilon^{(J_1 | ( I_1} F^{ I_2 \dots I_N) | J_2 \dots J_{N^\prime})} \,.
\es 
%
%

\subsection{Geometry of the comoving space}
We now proceed to lay out the geometric interpretation of the comoving formulation in terms of the \emph{comoving space}, defined as the set of codimension-one spacelike hypersurfaces that foliate spacetime and are transverse to the level sets of the scalar fields $\phi^I$, i.e., the particle worldlines. This comoving hypersurface picture will come in handy in later sections, where we introduce a complex scalar field living on it.

To start with, we identify a set of closed one-forms $e^I = d \phi^I$, which define the volume form on the comoving hypersurface  
\be \label{eq:volume}
 \omega = \frac{1}{2} \epsilon_{IJ} d\phi^I \wedge d\phi^J = \frac{1}{2} \epsilon_{IJ} e^I_\mu e^J_\nu dx^\mu \wedge dx^\nu \,,
\ee 
where we have denoted $e^I_\mu = \partial_\mu \phi^I$. This volume form is necessarily degenerate in $2+1$--dimensional spacetime, with its kernel spanned by the vector field 
\be \label{eq:vector}
v^\mu  =  \frac{1}{2 \sqrt{-g}} \epsilon^{\mu \nu \rho} \epsilon_{IJ} e^I_\nu e^J_\rho \,,
\ee 
where $v_\mu = (\star \omega)_\mu$ is the Hodge dual of the volume form Eq.~\eqref{eq:volume}. The worldlines of the particles are then identified with the integral curves of the vector field Eq.~\eqref{eq:vector}. It is useful to define the normalized velocity vector field $u^\mu = \frac{v^\mu}{b}$, with 
\be \label{eq:entropyDensity}
b = \sqrt{-v_\mu v^\mu} \,,
\ee 
which satisfies the relations 
\be \label{eq:velocity}
 u^\mu e^I_\mu = 0\,, \quad u^\mu u_\mu = -1 \,.
\ee  
The condition $ u^\mu e^I_\mu = 0$ implies that the $\phi^I$ coordinates remain constant along the fluid flow, thereby justifying the interpretation of $\phi^I$ as \textit{comoving coordinates}. Notice also that the quantities $u^\mu$ and $b$, which we will shortly identify as the fluid velocity and entropy density, are invariant under $\text{SDiff}(\mathbb R^2)$ transformations Eq.~\eqref{eq:diffeo}. Furthermore, the vector field Eq.~\eqref{eq:velocity} obeys a local conservation law\footnote{More generally, one has $\nabla_\mu \left( v^\mu f(\phi^I) \right) = 0$ for any function $f(\phi^I)$.} 
\be \label{eq:entropy}
\nabla_\mu v^\mu = 0 \rightarrow \nabla_\mu \left( b u^\mu \right) = 0\,, 
\ee 
reflecting the local conservation of entropy in the ideal fluid. 

From the one-forms $e^I_\mu$ we construct the comoving metric 
\be 
B^{IJ}=  g^{\mu \nu} e^I_\mu e^J_\nu\,, \quad \text{det}\, B^{IJ} = b^2\,,
\ee
with inverse $B_{IJ}$ satisfying $B_{JK} B^{KI} = \delta^I_J$. This metric measures the spatial distances on the comoving hypersurface
\be \label{eq:fluidMetric}
ds^2 = B_{IJ} d\phi^I \otimes d\phi^J\,.
\ee 
Pulling it back to spacetime 
\be 
ds^2 = B_{IJ}  e^I_\mu e^J_\nu dx^\mu \otimes dx^\nu\,, 
\ee 
where we have introduced the dual (tangent) basis vectors $e^\mu_I$, defined via
\be 
e^\mu_J e^I_\mu = \delta^I_J\,, \quad u_\mu e^\mu_J =0\,.
\ee 
With these ingredients, we can define the projector
\be \label{eq:projector}
\Pi^{\mu \nu} = B^{IJ}  e_I^\mu e_J^\nu = g^{\mu \nu} + u^\mu u^\nu\,,
\ee 
which projects onto the comoving hypersurface.
Throughout this work, we will use $B^{IJ}$ ($B_{IJ}$) to raise (lower) comoving indices whereas the spacetime indices are manipulated with the spacetime metric $g_{\mu \nu}$. 



\subsection{Effective action for uncharged fluids}

We now present a variational formulation for hydrodynamics of neutral fluids in terms of an action principle. To this end, we construct the most general action using the $\phi^I$ variables that is invariant under Poincaré transformations and APDs Eq.~\eqref{eq:APDs}. 
First, notice that any APD-invariant combination involving $\phi^I$ and their derivatives can be expressed in terms of the vector $v^\mu$ defined in Eq.~\eqref{eq:vector}. In particular, at lowest order in derivatives, the only Poincaré scalar is $b=\sqrt{-v_\mu v^\mu}$. 

Therefore, the leading order theory for the hydrodynamics of uncharged fluids can be expressed as \cite{Dubovsky:2005xd}
\be \label{eq:unchargedTheory}
S = \int d^3 x \sqrt{-g} \, F(b)\,,
\ee 
where $F(b)$ is an arbitrary function of the invariant scalar $b$. The variation of the action takes the form
\be 
\delta S = \int d^3 x \sqrt{-g} \Big[ -\frac{1}{2} T_{\mu \nu} \delta g^{\mu \nu} + \mathcal C_I \delta \phi^I \Big] \,.
\ee 
Demanding invariance under general coordinate transformations of spacetime coordinates gives the Ward identity\footnote{Under an infinitesimal diffeomorphism, $\delta_\xi g_{\mu \nu} = 2 \nabla_{(\mu} \xi_{\nu)}$ and $\delta_\xi \phi^I = \xi^\mu \partial_\mu \phi^I$.}
\be \label{eq:ward}
\nabla_\mu T^{\mu \nu} = \mathcal C_I e^{I \nu} \,,
\ee 
encoding covariant conservation of the stress-energy tensor whenever the $\phi^I$ fields are taken on-shell with respect to their equations of motion $\mathcal C_I = 0$. Conversely, $\nabla_\mu T^{\mu \nu} = 0$ implies $\mathcal C_I = 0$. Hence, the conservation of the stress-energy tensor is tantamount to imposing the equations of motion for the $\phi^I$ fields and fully determines their dynamics. 
Using the variational formulas (see Appendix \ref{app:geometricVariations}), from the action Eq.~\eqref{eq:unchargedTheory} we extract   
\be \label{eq:stressTensor}
T^{\mu \nu} = F g^{\mu \nu} - \sigma^\mu_I e^{I \nu}   \,, \quad \mathcal C_I = \nabla_\mu \sigma^\mu_I \,, 
\ee 
where $\sigma^\mu_I  = F_b b \, e^\mu_I$ with $F_b = \frac{dF}{db}$. Therefore, the dynamics of the theory Eq.~\eqref{eq:unchargedTheory} is fully captured by the conservation of the stress-energy tensor, $\nabla_\mu T^{\mu \nu} = 0$, with the constitutive relation 
\be 
 T_{\mu \nu} =-F_b b \, u_\mu u_\nu + g_{\mu \nu} \big(F  -F_b b \big)\,.
\ee 
Comparing with the relativistic Euler form $T^{\mu \nu} =\left( p + \epsilon \right) u^\mu u^\nu + p g^{\mu \nu}$, we identify the pressure and energy density 
\be \label{eq:pressure}
p = F -F_b b\,, \quad \epsilon = -F\,,
\ee 
as well as the temperature $T=-F_b$ and entropy density $s=b$ so that the thermodynamic relation $\epsilon + p = Ts$ is satisfied. Notice that the entropy current $S^\mu = b u^\mu$ is conserved off-shell $\nabla_\mu S^\mu = 0$ by virtue of Eq.~\eqref{eq:entropy}.

There are also conservation laws associated with $\text{SDiff}(\mathbb R^2)$ symmetry Eq.~\eqref{eq:APDs}, corresponding to vorticity conservation (see \cite{Cuomo:2024ekf}).

Finally, it is important to emphasize that the comoving formulation presented herein does not describe the most general hydrodynamic flows, but rather a subclass of homoentropic flows maintaining constant entropy per fluid element at all times. This is already apparent from the fact that the entropy density, Eq.~\eqref{eq:entropyDensity}, is fully determined by the mechanical displacements of the fluid elements, so that there is no room for genuine independent thermal transport. In particular, we have formulated the comoving theory in terms of only $d=2$ scalar fields $\phi^I$, whereas, in general, hydrodynamics admits $2+1$ independent variables\footnote{At the level of linearized ideal hydrodynamics, however, temperature fluctuations do not constitute an independent hydrodynamic mode and can be solved for in terms of velocity gradients via energy conservation.}. One way to remedy this is by introducing an ``internal clock'' variable, thereby enlarging the comoving manifold to $2+1$--dimensions parameterized by the coordinates $\phi^M$ with $M=0,1,2$ \cite{Nickel:2010pr}. However, such a choice leads to a generalized Euler relation whose physical interpretation is not clear at the moment \cite{Mancilla:2024spp}. Therefore, for the remainder of this work, we will stick with the comoving formulation $\phi^I$ with $I=1,2$.

\section{Charged fluids}\label{sec:charged}
Utilizing the comoving hypersurface interpretation, we propose a field-theoretic model for charged fluid dynamics that reproduces the nonlinear hydrodynamic equations of a perfect fluid. Contrary to previous formulations \cite{Dubovsky:2011sj}, the charge symmetry is realized linearly on the complex scalar field $\Phi$ residing on the comoving hypersurface. We provide a first-principles proof for the emergence of the chemical shift symmetry in charged fluids, corresponding to the comoving conservation of all multipole moments, and elucidate the resulting symmetry structure. Then, we formulate a symmetry-invariant action, derive the constitutive relations, and establish an explicit dictionary mapping the fields of the theory to hydrodynamic variables. We also discuss the equilibrium state, its symmetry-breaking pattern, and the spectrum of Nambu–Goldstone modes.

\subsection{Multipole symmetries on the comoving plane}
In this section we demonstrate that perfect charge-carrying fluids possess fractonic-like symmetries, which restrict the mobility of elementary charges on the comoving hypersurface.
In order to describe charged fluids, we introduce a complex scalar field $\Phi$ charged under the global $\text U(1)_{\text Q}$ symmetry. The $\Phi$ field transforms linearly under this symmetry 
\be \label{eq:charge}
\Phi \rightarrow e^{i \lambda_0} \Phi\,,
\ee 
where $\lambda_0$ is a constant parameter. 
By Noether's theorem, the symmetry Eq.~\eqref{eq:charge} implies the existence of a conserved current $\nabla_\mu J^\mu = 0$ such that the conserved charge can be expressed as
\be 
Q = \int d^2 x \sqrt{-g} J^0\,.
\ee 
In ordinary ideal fluids, charge is attached to the fluid elements, as these are the only dynamical degrees of freedom that can carry it. Consequently, the charge current takes the form $J^\mu = \rho u^\mu$, where $\rho$ is charge density.
Then, we observe the following identity 
\be 
\rho u^\mu \partial_\mu \left( \phi^{I_1} \cdots \phi^{I_N} \right) = 0\,,
\ee 
for any integer $N$ by virtue of Eq.~\eqref{eq:velocity}. It follows that there exists an infinite set of conserved currents 
\be 
\nabla_\mu J^{\mu I_1 \dots I_N} = 0\,, \quad J^{\mu I_1 \dots I_N} = J^\mu \phi^{I_1} \cdots \phi^{I_N} \,,
\ee 
together with the associated charges 
\be  \label{eq:multipoleMoments}
Q^{I_1 \dots I_N} = \int  d^2 x \sqrt{-g} J^0 \phi^{I_1} \cdots \phi^{I_N} \,.
\ee 
These charges reflect the conservation of the multipole moments of the charge density, familiar in the context of fractons \cite{Gromov:2018nbv}, with respect to the comoving hypersurface. Such conservation laws impose strong constraints on the dynamics of the elementary charges giving rise to fractonic behavior, wherein charges cannot move along the comoving hypersurface without violating one of the higher-moment conservation laws\footnote{Nevertheless, they can still be advected through spacetime by a fluid's motion.}.
A theory of charged fluids must respect the conservation of the multipole moments Eq.~\eqref{eq:multipoleMoments} and therefore be invariant under the symmetries generated by these charges,
\be 
\Phi \rightarrow e^{i \Lambda_{I_1 \dots I_N} \phi^{I_1} \cdots \phi^{I_N}} \Phi\,, 
\ee 
where $\Lambda_{I_1 \dots I_N}$ is a constant symmetric tensor of parameters. 
Since this must hold for any $N$, we conclude that in ordinary fluids the charge symmetry Eq.~\eqref{eq:charge} must be promoted to the \textit{chemical shift} symmetry\footnote{The chemical shift symmetry was first introduced in \cite{Dubovsky:2011sj} to construct an effective field theory for hydrodynamics. In our formulation, however, it is realized linearly on the complex scalar field and emerges naturally as a consequence of Noether’s theorem.}
\be \label{eq:chemicalShift}
\Phi \rightarrow e^{i \Lambda(\phi^I)} \Phi\,,
\ee 
where $\Lambda(\phi^I)$ is an arbitrary function that admits a polynomial expansion
\be 
\Lambda(\phi^I) = \sum_{N=0}^\infty  \Lambda_{I_1 \dots I_N} \phi^{I_1} \cdots \phi^{I_N}\,.
\ee 
The component of degree-$N$ generates a polynomial shift symmetry
\be \label{eq:multipole}
\delta_{\Lambda_N} \Phi = i \Lambda_{I_1 \dots I_N} \phi^{I_1} \cdots \phi^{I_N} \Phi \,. 
\ee 
The multipole shifts are abelian but have nontrivial commutation relations with the APDs Eq. \eqref{eq:SigmaN}. In particular, we find 
\be 
[\delta_{\Sigma_N}, \delta_{\Lambda_{N^\prime}}] \Phi = i N^\prime \epsilon^{I_1 J_1} \Lambda_{I_1 \dots I_{N^\prime}}      \Sigma_{J_1 J_2 \dots J_N}  \phi^{I_2} \cdots \phi^{I_{N^\prime}} \phi^{J_2} \cdots \phi^{J_N} \Phi \,.
\ee 
Therefore, the symmetry variations $\{\delta_{\Sigma_N}, \delta_{\Lambda_N}\}$ furnish a closed algebra 
\be \label{eq:algebraVariations}
[\delta_{\Sigma_N}, \delta_{\Lambda_{N^\prime}}]  = \delta_{\Lambda_{N^{\prime \prime}}}  \,, \quad [\delta_{\Sigma_N}, \delta_{\Sigma_{N^\prime}}] = \delta_{\Sigma_{N^{\prime \prime}}} \,,
\ee 
where $N^{\prime \prime} = N + N^\prime - 2$ and
\bs 
\Lambda_{I_1 \dots I_{N^{\prime \prime}}} &= N^\prime \epsilon^{I J} \Lambda_{I I_1 \dots I_{N^\prime-1}}      \Sigma_{J I_{N^\prime} \dots I_{N^{\prime \prime}} }\,, \\
\Sigma^{\prime \prime}_{I_1 \dots I_{N^{\prime \prime}}} &= N^{\prime \prime} \epsilon^{I J}  \Sigma^\prime_{ I I_1 \dots I_{N^\prime-1} } \Sigma_{J I_{N^\prime}  \dots I_{N^{\prime \prime}}}\,.
\es 
%
Introducing a polynomial basis of the Lie algebra generators
\be \label{eq:generatorsP}
\delta_{\Lambda_N} = i \Lambda_{I_1 \dots I_N} P^{I_1 \dots I_N}\,, \quad \delta_{\Sigma_N} = \Sigma_{I_1 \dots I_N}  F^{I_1 \dots I_N}\,, 
\ee 
where $P^{I_1 \dots I_N}  = \phi^{I_1} \cdots \phi^{I_N}$ is a homogeneous polynomial of degree $N$ and $F^{I_1 \dots I_N}$ is defined in Eq.~\eqref{eq:generatorsF}. 
Then, the full symmetry algebra is specified with the following commutation relations 
\bs 
[F^{I_1 \dots I_N}, P^{J_1 \dots J_{N^\prime}}] &= N^\prime \epsilon^{(J_1|  (I_1} P^{I_2 \dots I_N) | J_2 \dots J_{N^\prime})}\,, \\
[F^{I_1 \dots I_N}, F^{J_1 \dots J_{N^\prime}}] &= N^{\prime \prime} \epsilon^{(J_1 | ( I_1} F^{ I_2 \dots I_N) | J_2 \dots J_{N^\prime})} \,.
\es 

\subsection{The model}
Having established the appropriate degrees of freedom for charged fluids and their symmetry structure, we are now equipped to write down an action governing their hydrodynamics. For this purpose, we consider the following action\footnote{The theory Eq.~\eqref{eq:normalAction} can be understood as a UV completion of the effective hydrodynamic theory constructed in \cite{Dubovsky:2011sj}.}
\be  \label{eq:normalAction}
S = \int d^3 x \sqrt{-g} \Big[ \frac{i}{2}\big( \Phi^\dag D_0 \Phi - \Phi D_0 \Phi^\dag \big) - E(|\Phi|, b) \Big] \,,
\ee
where $D_0 \Phi = u^\mu (\partial_\mu - i A_\mu) \Phi$ and $E(|\Phi|, b)$ is an arbitrary function. Throughout this manuscript, we assume that $E(|\Phi|, b)$ is factorizable\footnote{This choice corresponds to a subclass of fluids that admit a factorizable equation of state.}, $E(|\Phi|, b) = V(|\Phi|) - F(b)$, and choose the potential $V(|\Phi|) = \frac{\lambda}{2} |\Phi|^4$, describing short-range repulsive interactions, while relegating the study of more general scenarios to future work\footnote{In full generality, the action can also exhibit arbitrary dependence on the symmetry-invariant combination $Y \equiv i \left( \Phi^\dag D_0 \Phi - \Phi D_0 \Phi^\dag \right)$.}. 

It is straightforward to verify that the theory Eq. \eqref{eq:normalAction} is invariant under Poincaré transformations, APDs \eqref{eq:diffeo}, and chemical shifts \eqref{eq:chemicalShift}. In total, the symmetry group is
\be 
G = \text{ISO}(2,1) \times \left( \text{SDiff}(\mathbb R^2) \ltimes \text{CShift}(\infty) \right)\,.
\ee 
Furthermore, we have coupled the theory to the metric and gauge field rendering it covariant under general transformations of spacetime coordinates and local $\text U(1)_{\text Q}$ transformations.
Despite being Poincaré invariant the action \eqref{eq:normalAction} contains only a single time derivative, reminiscent of nonrelativistic field theories. As a consequence, $\Phi$ and $\Phi^\dagger$ are not independent, so they account for only a single propagating degree of freedom. With this in mind, it is convenient to employ the polar decomposition $\Phi = \sqrt{\rho} e^{i \psi}$ such that the action \eqref{eq:normalAction} takes the following form 
\be \label{eq:normalActionPolar}
S = \int d^3 x \sqrt{-g} \Big[ -\rho D_0 \psi - \frac{\lambda}{2} \rho^2 + F(b) \Big] \,,
\ee 
where $D_0 \psi = u^\mu \left( \partial_\mu \psi - A_\mu \right)$. 
Indeed, we see that $\rho$ is an auxiliary variable and can be eliminated from its equations of motion 
\be \label{eq:conditionRhoNormal}
\rho = -\frac{1}{\lambda} D_0 \psi\,.
\ee 
In the following, we find it most convenient to work with the action in the polar form Eq. \eqref{eq:normalActionPolar} and invoke \eqref{eq:conditionRhoNormal} when needed.

\subsection{Relativistic Euler equations}
In this section, we analyse the equations of motion and conservation laws of the action \eqref{eq:normalActionPolar}. We find that the resulting dynamics is governed by the relativistic Euler equations, as expected for ideal charged fluids. This confirms that the model \eqref{eq:normalAction} describes an ideal relativistic fluid with a conserved charge.

Let us begin by establishing the general structure of the equations of motion. A variation of the action Eq. \eqref{eq:normalActionPolar} with respect to the external sources $g^{\mu \nu}$ and $A_\mu$, as well as its dynamical fields $\phi^I$ and $\psi$, takes the following form
\be \label{eq:variationsAction}
\delta S = \int d^3 x \sqrt{-g} \Big[ -\frac{1}{2} T_{\mu \nu} \delta g^{\mu \nu} + J^\mu \delta A_\mu + \mathcal C_I \delta \phi^I + \mathcal K \delta \psi  \Big] \,.
\ee 
The classical dynamics of the system is then determined by the respective equations of motion, $\mathcal C_I = 0$ and $\mathcal K = 0$.
Under infinitesimal diffeomorphism and local $\text U(1)_{\text Q}$ transformations parameterized by $\chi = (\xi^\mu, \alpha)$, the fields transform as 
\bs 
\delta_\chi g^{\mu \nu} &= \mathcal L_\xi g^{\mu \nu} =  \nabla^\mu \xi^\nu + \nabla^\nu \xi^\mu \,, \\
\delta_\chi A_\mu &= \mathcal L_\xi A_\mu + \partial_\mu \alpha = \xi^\nu \partial_\nu A_\mu + A_\nu \partial_\mu \xi^\nu + \partial_\mu \alpha\,, \\ 
\delta_\chi \phi^I &=  \mathcal L_\xi \phi^I = \xi^\mu e^I_\mu \,, \\ 
\delta_\chi \psi &= \mathcal L_\xi \psi +  \alpha  = \xi^\mu \partial_\mu \psi +  \alpha \,.
\es 
Demanding invariance under $\delta_\chi$ transformations yields the following Ward identities 
\bs \label{eq:wardCharged}
\nabla_\mu T^{\mu \nu} &= F^{\nu \mu} J_\mu + \mathcal C_I e^{I \nu} + \mathcal K  \partial^\nu \psi \,, \\
\nabla_\mu J^\mu &=  \mathcal K \,,
\es 
where we have defined the field strength tensor of the $\text U(1)_{\text Q}$ gauge field $A_\mu$ as
\be 
F_{\mu \nu} \equiv  \partial_\mu A_\nu - \partial_\nu A_\mu \,. 
\ee 
These identities express the covariant conservation of the energy-momentum tensor\footnote{Including the source term due to the Lorentz force.} and of the charge current, whenever $\phi^I$ and $\psi$ are taken on-shell, i.e., $\mathcal C_I = 0$ and $\mathcal K = 0$. 
On the other hand, imposing the conservation equations 
\be \label{eq:conserved}
\nabla_\mu T^{\mu \nu} = F^{\nu \nu} J_\mu\,, \quad 
\nabla_\mu J^\mu = 0\,,
\ee 
implies the simultaneous vanishing of the source terms  $\mathcal C_I$ and $\mathcal K$, thereby putting the fields on-shell. Thus, we conclude that the full dynamics of the system is equivalently governed by the conservation equations \eqref{eq:conserved}.

Our strategy is to compute the currents $T^{\mu \nu}$ and $J^\mu$ from the action Eq.~\eqref{eq:normalActionPolar} and recast them in the form of hydrodynamic equations. After doing so, we establish an explicit mapping between the dynamical fields of the theory and the hydrodynamic variables. 
Using the variational formulae collected in the Appendix \ref{app:geometricVariations} we find
\be \begin{split} \label{eq:constitutiveNormal}
J^\mu &= \rho u^\mu\,, \\
T^{\mu \nu} &= \big( \lambda \rho^2  - F_b b\big)  u^\mu u^\nu + \big(\mathcal L - F_b b\big) g^{\mu \nu}\,,
\end{split}
\ee
where in the second line we have also used \eqref{eq:conditionRhoNormal}. Comparing with the constitutive relations of a perfect fluid,
\be
T^{\mu \nu} = (\epsilon + p) u^\mu u^\nu + p g^{\mu \nu}\,, \quad j^\mu = \rho u^\mu\,,
\ee 
we identify $\rho$ as charge density whereas the expressions for pressure and energy density read
\be 
p = \frac{\lambda}{2} \rho^2 + F - F_b b\,, \quad \epsilon =  \frac{\lambda}{2} \rho^2 - F\,.
\ee 
Moreover, utilizing the thermodynamic identity $\epsilon + p = \rho \mu + Ts$ we also establish the following relations 
\be \label{eq:thermRelations}
\mu = \lambda \rho \,, \quad T = - F_b\,, \quad s = b\,.
\ee 
We see that $\mu \equiv \mu(\rho)$ and $T \equiv T(s)$, indicating a fluid with a factorizable equation of state, which is a consequence of the assumption $G(|\Phi|, b) = V(|\Phi|) - F(b)$.
Moreover, notice that we have expressed all hydrodynamic variables $(\mu, T, u^\mu)$ in terms of the fields $(\phi^I, \psi)$. The proposed action Eq.~\eqref{eq:normalAction} thus reproduces the relativistic Euler equations and encodes hydrodynamics of charged fluids more economically, using only $(d+1)$ independent dynamical fields instead of $(d+2)$ hydrodynamic variables.

\subsection{Thermodynamic equilibrium and symmetry breaking}\label{sec:equilibrium}
We now consider a finite density state at thermodynamic equilibrium placed in flat Minkowski spacetime $g_{\mu \nu} = \eta_{\mu \nu}$ and discuss the associated symmetry breaking pattern. 
For a homogeneous equilibrium we can always set 
\be \label{eq:equilibrium}
\langle \phi^I \rangle_{eq} = \sqrt{s_0} \,  \delta^I_i x^i\,,
\ee 
where $s_0$ is a numerical prefactor whose role we elucidate shortly. This configuration is achieved by exploiting the $\text{SDiff}(\mathbb R^2)$ symmetry \eqref{eq:diffeo} to align the fluid coordinates with the spatial coordinates of spacetime\footnote{Hence the arbitrary prefactor, which cannot be removed by an APD.}. 
In doing so, we spontaneously break the $\text{SDiff}(\mathbb R^2)$ and spacetime Poincaré symmetry down to a diagonal unbroken subgroup consisting of simultaneous spatial translations $P_i$ and fluid shifts $F^I$, generated by the combination 
\be \label{eq:unbroken}
\bar P_i = P_i + \sqrt{s_0} \delta^I_i \epsilon_{IJ} F^J \,.
\ee 
Indeed, it is straightforward to check that the action of the unbroken translation with parameter $c^i$ send 
\be 
x^i \rightarrow x^i + c^i\,, \quad \phi^I \rightarrow \phi^I + \sqrt{s_0} \delta^I_i c^i\,, 
\ee 
and therefore leaves the equilibrium configuration Eq.~\eqref{eq:equilibrium} intact. There is also a notion of unbroken rotational symmetry associated to the combination 
\be 
\bar J = J + \sqrt{s_0} \delta_{IJ} F^{IJ}\,, 
\ee 
where $J = - \epsilon_{ij} x_i \partial_j$ generates infinitesimal spacetime rotations.
Moreover, notice that \eqref{eq:equilibrium} describes a static equilibrium, $u^{\mu} = (1,0)$, which can always be obtained by boosting to the rest frame of the system, signalling the spontaneous breaking of the Lorentz boost symmetry.

Furthermore, we place a system at a finite chemical potential $\mu>0$ by switching on a background gauge field $A_\mu = ( \mu_0,0)$. Consequently, the effective potential evaluated on the equilibrium fluid configuration \eqref{eq:equilibrium} takes the form of the Mexican hat potential
\be \label{eq:effectivePotential}
V(|\Phi|) = - \mu_0 |\Phi|^2 + \frac{\lambda}{2} |\Phi|^4\,. 
\ee 
Therefore, in a homogenous finite density state the $\Phi$ field develops a vacuum expectation value $\langle \Phi \rangle  = \sqrt{\rho_0} $ with $\rho_0 = \frac{\mu_0}{\lambda}$, breaking spontaneously the chemical shift symmetry \eqref{eq:chemicalShift}.  
In summary, the symmetry breaking pattern is 
\be \label{eq:SSBPattern}
 \text{ISO}(2,1)  \times \left( \text{SDiff}(\mathbb R^2) \ltimes \text{CShift}(\infty) \right)  \rightarrow \left( \text{SO}(2) \ltimes \mathbb R^2 \right) \times \mathbb R\,.
\ee

\subsection{Linearized hydrodynamics}
In this section, we derive the effective theory that governs the long-wavelength dynamics around a homogeneous thermal equilibrium state at finite density.
With this in mind, we consider fluctuations around the ground state 
\be \label{eq:fluctations}
\Phi = \sqrt{\rho_0+\delta \rho}\, e^{i \varphi}\,, \quad \phi^I = \sqrt{s_0} \, \delta^I_i \big(x^i + \pi^i\big)\,,
\ee
where we have introduced the Nambu-Goldstone fields $\varphi$ and $\pi^i$ that provide a nonlinear realization of the symmetry breaking pattern Eq.~\eqref{eq:SSBPattern}.
Plugging~\eqref{eq:fluctations} into Eq.~\eqref{eq:normalActionPolar} we find that the effective Lagrangian truncated at second order in fluctuations is
\be \label{eq:normalFluidEFT}
\mathcal L = - \rho_0 \partial_t \varphi +\frac{1}{2 \lambda} (\partial_t \varphi)^2 + \rho_0 \partial_t \pi^i \partial_i \varphi   + \frac{w_0}{2} (\partial_t \pi^i)^2 +\frac{F_{bb} s_0^2}{2} (\partial_i \pi^i)^2  \,,
\ee 
where we have dropped terms that are higher than quadratic in fluctuations, integrated out $\delta \rho = - \frac{1}{\lambda} \partial_t \varphi$, and defined enthalpy density $w_0 = \rho_0 \mu_0 - F_b s_0$ (see Appendix \ref{app:linearized} for details). 
The equations of motion for the effective Lagrangian Eq.~\eqref{eq:normalFluidEFT} take the form of linearized Euler equations for the relativistic fluid 
\be \begin{split}
\partial_t \delta \rho + \rho_0 \partial_i \delta u^i &= 0 \,, \\
\left(\epsilon_0 + p_0 \right) \partial_t \delta u^i + \partial_i \delta p &=0 \,,
\end{split}
\ee 
where $\delta p = s_0 \delta T + \rho_0 \delta \mu$ with
\be \label{eq:thermDefs}
\delta T = -F_{bb}  \delta s\,, \quad \delta s = s_0 \partial_i \pi^i\,, \quad \delta \mu =  \lambda \delta \rho \,, \quad \delta u^i = - \partial_t \pi^i\,.
\ee 
Recalling the off-shell identity Eq.~\eqref{eq:entropy},
\be 
\partial_\mu v^\mu = \partial_\mu \left( b u^\mu \right) = 0\,,
\ee
which expresses local conservation of the entropy current in ideal fluids, we can recast the dynamical equations as 
\be \begin{split}
\partial_t \delta \rho + \rho_0 \partial_i \delta u^i &= 0 \,, \\
\left(\epsilon_0 + p_0 \right) \partial_t \delta u^i - F_{bb} s_0 \partial_i \delta s + \mu_0 \partial_i \delta \rho &=0 \,, \\
\partial_t \delta s + s_0 \partial_i \delta u^i &=0\,.
\end{split}
\ee 
These equations admit a wavelike propagating solution corresponding to the longitudinal sound mode with a linear dispersion 
\be \label{eq:sound}
\omega = \pm c_s^2 k\,, \quad c_s^2 = \frac{ \rho_0 \mu_0 - F_{bb} s_0^2}{w_0}\,.
\ee 
Using \eqref{eq:thermDefs}, the speed of sound can be recast in the conventional form \cite{landau1987fluid}
\be 
c_s^2 = \frac{\rho_0}{w_0}   \left(\frac{\partial p}{\partial s} \right)_\rho + \frac{s_0}{w_0}  \left(\frac{\partial p}{\partial \rho} \right)_s = \left( \frac{\partial p}{\partial \epsilon} \right)_{S,N}\,,
\ee
where the second equality follows after applying a number of thermodynamic identities (see \cite{Herzog:2008he}).

In total, there are three Goldstone fields, but only one propagating mode associated with the longitudinal component $\pi_{||}$. The remaining two modes admit a trivial solution $\omega = 0$, reflecting their fractonic nature. This can be attributed to the presence of restrictive $\text{SDiff}(\mathbb R^2)$ and $\text{CShift}(\infty)$ symmetries.
In particular, under an infinitesimal $\text{SDiff}(\mathbb R^2)$ transformation the $\pi^i$ Goldstone shifts as 
\be 
\delta_\Sigma \pi^i = \frac{1}{\sqrt{s_0}} \epsilon^{ij} \partial_j \Sigma(\sqrt{s_0} x) + \mathcal{O}(\pi) \,. 
\ee 
Hence, $\text{SDiff}(\mathbb R^2)$ acts on the transverse component $\pi_\perp$, forbidding the kinetic term for this mode (see also \cite{Endlich:2010hf}). 
Similarly, the $\varphi$ Goldstone realizes nonlinearly not only the $\text U(1)_Q$ symmetry but rather the full chemical shift symmetry 
\be 
\delta_\Lambda \varphi = f(\sqrt{s_0} x) + \mathcal{O}(\pi)\,,  
\ee 
disallowing a kinetic term for $\varphi$. In the superfluid phase, the chemical shift symmetry $\text{CShift}(\infty)$ is relaxed down to $\text U(1)_Q$ and the $\varphi$ Goldstone acquires a kinetic term, giving rise to the second sound phenomenon.  

\section{Superfluids}\label{sec:superfluid}
We now move our attention to the superfluid phase and introduce an action principle governing the hydrodynamics of the two-fluid model developed by Tisza and Landau in the context of superfluid helium-4 \cite{Tisza:1938ydz,Landau:1941lul}. While complex scalar field theories for zero-temperature superfluids are well established, their finite-temperature counterparts are lacking. An EFT for finite-temperature superfluids, formulated in terms of a set of Goldstone fields, was developed by Alberto Nicolis in Ref.~\cite{Nicolis:2011cs}. The model introduced here can be regarded as a UV completion of \cite{Nicolis:2011cs}. 

After postulating an action, we verify that our model reproduces the nonlinear superfluid hydrodynamic equations in the infrared. We then elucidate the physical content of the theory by studying the dynamics of long-wavelength fluctuations around an equilibrium state. In our formulation, certain effective coefficients can be traced back to their microscopic origin in the action. For instance, the coupling between the superfluid and normal components is fixed by the equilibrium charge density.

\subsection{The two-fluid model}
In finite-temperature superfluids, charge can be transported by both the normal and superfluid components, in accordance with the two-fluid model. In particular, the charge current is no longer proportional to the velocity of the thermal component $J^\mu \neq  \rho u^\mu$. Therefore, in the superfluid phase we drop the fractonic symmetry Eq.~\eqref{eq:chemicalShift}, retaining only the constant shifts Eq.~\eqref{eq:charge} so that the symmetry group is 
\be 
G = \text{ISO}(2,1)  \times  \text{SDiff}(\mathbb R^2) \times \text U(1)_{\text Q}\,.
\ee 
With this in mind, we propose the following theory
\be \label{eq:superfluidTheory}
S = \int d^3 x \sqrt{-g} \Big[ \frac{i}{2}\big( \Phi^\dag D_0 \Phi -\Phi D_0 \Phi^\dag \big)  -\frac{1}{2 m}  B^{IJ} D_I \Phi^\dag D_J \Phi - V(|\Phi|) + F(b)\Big] \,,
\ee 
where $D_I = e^\mu_I D_\mu = e^\mu_I  \left( \partial_\mu - i A_\mu \right)$. 
The second term allows charge transport in directions transverse to the fluid velocity, which is forbidden in the normal phase by the chemical shift symmetry Eq.~\eqref{eq:chemicalShift}. While charges can now redistribute freely within the comoving plane, they remain confined to it. Using Eq.~\eqref{eq:projector} we can express the Lagrangian in terms of the projector 
\be
\mathcal L =  \frac{i}{2}\big( \Phi^\dag D_0 \Phi -\Phi D_0 \Phi^\dag \big)  - \frac{1}{2m} \Pi^{\mu \nu} D_\mu \Phi^\dagger  D_\nu \Phi - V(|\Phi|) + F(b)
\,. 
\ee  
Notice that despite its unusual form with a single time derivative the superfluid action is still exactly invariant under Poincaré transformations. 
Employing polar decomposition $\Phi = \sqrt{\rho} e^{i \psi}$ the superfluid Lagrangian is
\be 
\mathcal L = -\rho D_0 \psi  - \frac{\lambda}{2} \rho^2  - \frac{\rho}{2m}    \Pi^{\mu \nu} D_\mu \psi  D_\nu \psi -\frac{1}{8m\rho}  \Pi^{\mu \nu} \partial_\mu \rho  \partial_\nu \rho + F(b) \,.
\ee 
The equation of motion for $\rho$ yields 
\be \label{eq:rhoSuperfluid}
\rho = - \frac{1}{\lambda} \left( D_0 \psi + \frac{1}{2m} \Pi^{\mu \nu} D_\mu \psi  D_\nu \psi \right) + \dots \,,
\ee 
where the dots represent subleading corrections coming from the quantum pressure term, which can be dropped in the hydrodynamic regime. In principle, we could substitute the expression for $\rho$ back into the superfluid Lagrangian 
\be 
\mathcal L =  \frac{1}{2\lambda} \left( D_0 \psi + \frac{1}{2m} \Pi^{\mu \nu} D_\mu \psi  D_\nu \psi \right)^2 + F(b) \,,
\ee 
where we have neglected the higher-derivative contributions from the quantum pressure. 
However, we find it more convenient to work instead with the theory
\be \label{eq:superfluid}
 S = \int d^3 x \sqrt{-g} \Big[ -\rho D_0 \psi  - \frac{\lambda}{2} \rho^2  - \frac{\rho}{2m}    \Pi^{\mu \nu} D_\mu \psi  D_\nu \psi  + F(b) \Big] \,,
\ee 
and keep in mind the relation \eqref{eq:rhoSuperfluid}.

\subsection{Constitutive relations}
In this section, we analyse the dynamical equations of the superfluid model and establish a dictionary between the scalar fields and hydrodynamic variables. By doing so, we verify that theory \eqref{eq:superfluid} correctly encodes the nonlinear hydrodynamic equations for finite temperature superfluids in accordance with Landau's two fluid model.

The variation of the superfluid action is given by Eq.~\eqref{eq:variationsAction} and the dynamics of the system is encoded in the conservation equations Eq.~\eqref{eq:conserved}. From the superfluid action Eq.~\eqref{eq:superfluid} we can read off the constitutive relations for the hydrodynamic currents. 
For the charge current we find
\be \label{eq:conservationSuperfluid}
 J^\mu  = \rho u^\mu + \frac{\rho}{m}  \zeta^\mu \,,
\ee 
where we have defined the transverse component of the superfluid velocity 
\be \label{eq:superfluidVelocity}
\zeta^\mu =  \Pi^{\mu \nu} D_\nu \psi\,.
\ee 
Notice that $J^\mu$ is no longer proportional to $u^\mu$, indicating that in superfluids, charge can flow transverse to the fluid velocity.

Varying the action Eq.~\eqref{eq:superfluid} with respect to the metric, using the variational formulas listed in Appendix~\ref{app:geometricVariations}, we obtain the following constitutive relation for the stress-energy tensor 
\be \label{eq:stressTensorSuperfluid}
T_{\mu \nu} =  \frac{\rho}{m}    D_\mu \psi  D_\nu \psi +  \left( -\frac{\rho}{m} (D_0 \psi)^2  -\rho D_0 \psi - F_b b \right)  u_\mu u_\nu + \left( \mathcal L-F_b b \right) g_{\mu \nu}\,.
\ee
It is convenient to rewrite the stress-energy tensor in terms of the transverse superfluid velocity defined in Eq.~\eqref{eq:superfluidVelocity}, giving 
\be
T_{\mu \nu} = \frac{\rho}{m} \zeta_\mu \zeta_\nu -\frac{2 \rho}{m} D_0 \psi \zeta_{(\mu} u_{\nu)}  +  \left(  -\rho D_0 \psi - F_b b \right)  u_\mu u_\nu + \left( \mathcal L-F_b b \right) g_{\mu \nu}\,.
\ee
Matching to the superfluid constitutive relations \cite{Arean:2023nnn} we establish the following identifications
\be \begin{split} \label{eq:superfluidPressure}
p &=  -\frac{\rho}{2} D_0 \psi   - \frac{\rho}{4m} \zeta^2   + F - F_b b\,, \\
\epsilon &=  -\frac{\rho}{2} D_0 \psi + \frac{\rho}{4m} \zeta^2   - F\,, \\
\mu &= -D_0 \psi \,, \\
n_s &= \frac{\rho}{m} \mu\,.
\end{split}
\ee 
Invoking the thermodynamic relation $\epsilon + p = \rho \mu + Ts $ we also identify $T = - F_b$ and $s = b$. Using Eq.~\eqref{eq:rhoSuperfluid} it is then possible to verify the thermodynamic identity 
\be
dp = s dT + n d\mu - \frac{n_s}{2 \mu} d \zeta^2 \,, 
\ee
in full agreement with \cite{Arean:2023nnn}.

\subsection{Linear response and second sound}
The discussion of the equilibrium state parallels that of ordinary fluids laid out in Sec.~\ref{sec:equilibrium}, except that here the symmetry breaking pattern is modified to
\be 
 \text{ISO}(2,1)  \times  \text{SDiff}(\mathbb R^2) \times \text U(1)_Q  \rightarrow \left( \text{SO}(2) \ltimes \mathbb R^2 \right) \times \mathbb R\,.
\ee 
As before, we restrict to flat Minkowski spacetime $g_{\mu \nu} = \eta_{\mu \nu}$ at finite chemical potential $A_\mu = (\mu_0,0)$. Expanding the superfluid action Eq.~\eqref{eq:superfluid} around the equilibrium state Eq.~\eqref{eq:fluctations}, and proceeding analogously to the derivation of Eq.~\eqref{eq:normalFluidEFT}, we arrive at the following effective theory  
\be \label{eq:superfluidEffective}
\mathcal L  = \frac{1}{2 \lambda} (\partial_t \varphi)^2   + \rho_0 \partial_t \pi^i \partial_i \varphi - \frac{\rho_0 }{2m}  (\partial_i  \varphi)^2 + \frac{w_0}{2} (\partial_t \pi^i)^2 +\frac{F_{bb} s_0^2}{2} (\partial_i \pi^i)^2 \,.
\ee 
The only distinction from the EFT Eq.~\eqref{eq:normalFluidEFT} for the normal fluid is the presence of a kinetic term for $\varphi$, which is allowed due to the absence of the chemical shift symmetry in the superfluid phase.

Eq.~\eqref{eq:superfluidEffective} bears close resemblance to the quadratic theory Eq.~(17) of Ref.~\cite{Nicolis:2011cs}. However, Eq.~\eqref{eq:superfluidEffective} contains fewer free parameters and provides a more microscopic interpretation to some of them. For example, the coefficient in front of $(\partial_t \varphi)^2$ is fixed by the strength of interaction $\lambda$, that of $\partial_t \pi^i \partial_i \varphi$ is set by equilibrium charge density $\rho_0$, and that of $(\partial_i  \varphi)^2$ by the ratio $\frac{\rho_0}{m}$. 

Furthermore, Eq.~\eqref{eq:superfluidEffective} differs structurally from \cite{Nicolis:2011cs} in that it contains a $\partial_t \pi^i \partial_i \varphi$ coupling, instead of the $\partial_i \pi^i \partial_t \varphi$ term. While these terms are equivalent for smooth configurations up to a total derivative, they differ in the presence of singularities associated with superfluid vortices, i.e. when $\epsilon^{\mu \nu \rho} \partial_\nu \partial_\rho \varphi \neq 0$.
To elucidate the role of superfluid vortices let us analyse the equation of motion for $\pi^i$,
\be 
\partial_t \left( w_0 \partial_t \pi^i - \rho_0 \partial_i \varphi \right) + F_{bb} s_0^2 \partial_i  \partial_j \pi^j = 0\,.
\ee 
Projecting the equation along the transverse direction $\epsilon_{ij} \partial_j$ and assuming static vortices, i.e. $[\partial_t, \partial_i] \varphi = 0$ but $[\partial_i, \partial_j] \varphi \neq 0$, we arrive at 
\be \label{eq:vorticity}
\partial_t \left( w_0 \epsilon^{ij} \partial_j \partial_t \pi^i + \rho_0 \epsilon^{ij} \partial_i \partial_j \varphi \right) = 0\,.
\ee 
Since $\partial_t \pi^i$ represents the fluctuation in the velocity of the normal component (see Eq.~\eqref{eq:thermDefs}), we recognize 
\be
\delta \omega = \epsilon^{ij} \partial_j \partial_t \pi^i =  \epsilon^{ij}\partial_i \delta u^j\,,
\ee 
as the linearized vorticity of the normal component. Then, Eq.~\eqref{eq:vorticity} admits a clear physical interpretation as a conservation equation for the total vorticity  
\be
\partial_t \left( w_0 \delta \omega + \rho_0 \epsilon^{ij} \partial_i \partial_j \varphi \right) = 0\,, 
\ee 
where the first term represents the contribution from the normal component and the second from superfluid vortices. In particular, a fluctuation of vorticity in the normal component can nucleate a superfluid vortex, such that the total vorticity remains conserved.

We now turn to the analysis of the excitation spectrum. For this purpose, we assume smooth configurations and preform a Fourier transformation of the Lagrangian Eq.~\eqref{eq:superfluidEffective} 
\be
S =\frac{1}{2} \int \frac{ d^2 k d\omega }{(2\pi)^{3/2}}   \begin{pmatrix}
    \tilde \varphi(k,\omega) \\
    \tilde \pi_{||}(k,\omega) \\
    \tilde \pi_{\perp}(k,\omega) 
\end{pmatrix}^T \begin{pmatrix}
    \frac{1}{\lambda} \omega^2 - \frac{\rho_0}{m}  k^2 & \rho_0 \omega k & 0\\
    \rho_0 \omega k & w_0 \omega^2 + F_{bb} s_0^2 k^2 & 0\\
    0  & 0 & w_0 \omega^2
\end{pmatrix}\begin{pmatrix}
   \tilde \varphi(-k,-\omega) \\
   \tilde \pi_{||}(-k,-\omega) \\
   \tilde \pi_{\perp}(-k,-\omega) 
\end{pmatrix}\,.
\ee 
After solving the characteristic equation, we verify the existence of two propagating modes with a linear dispersion relation
\bs 
\omega_1 = \pm v_1 k\,, \\
\omega_2 = \pm v_2 k\,, 
\es 
corresponding to first and second sound with the velocities given by 
\bs \label{eq:secondSound}
v_1^2 &= \frac{c_s^2}{2} + \frac{\lambda \rho_0}{2 m} + \frac{1}{m} \sqrt{(m c_s^2 + \lambda \rho_0)^2 + 4 m s_0^2 \rho_0 \lambda F_{bb} }\,, \\
v_2^2 &= \frac{c_s^2}{2} + \frac{\lambda \rho_0}{2 m} - \frac{1}{m} \sqrt{(m c_s^2 + \lambda \rho_0)^2 + 4 m s_0^2 \rho_0 \lambda F_{bb} }\,,
\es  
where $c_s^2$ is defined as in Eq.~\eqref{eq:sound}. On the other hand, the transverse component $\pi_\perp$ is still non-propagating, $\omega = 0$, as a result of the $\text{SDiff}(\mathbb R^2)$ symmetry. 
In order to better understand the expressions \eqref{eq:secondSound} it is helpful to perform an expansion. First, we consider an expansion for the large values of $m$, giving
\bs 
v_1^2 &= c_s^2 + \mathcal{O}(\frac{1}{m})\,, \\
v_2^2 &=  \mathcal{O}(\frac{1}{m}) \,.
\es
We see that in the limit $m \rightarrow \infty$, the first mode reduces to the ordinary sound mode of normal fluids Eq.~\eqref{eq:sound}, whereas the second becomes non-propagating. 
Expanding to first order in $\lambda$ at fixed density ($\mu_0  = \lambda \rho_0$), we find 
\bs 
v_1^2 &= - \lambda \frac{s_0^2 F_{bb}}{\rho_0^2} + \mathcal{O}(\lambda^2)\,, \\
v_2^2 &= \lambda \frac{\rho_0}{m} + \mathcal{O}(\lambda^2) \,.
\es
One of the modes depends only on the entropy (thermal) sector, while the second one, controlled by the charge density, reproduces the familiar dispersion relation of the Bogoliubov phonon \cite{pitaevskii2016bose}. Therefore, in the small $\lambda$ limit the two solutions decouple into a charge-dominated Bogoliubov phonon and a thermal (entropy-dominated) mode.

\section{Fracton fluids}\label{sec:fracton}
So far, we have considered ordinary charged fluids, whose dynamics is constrained by the chemical shift symmetry Eq.~\eqref{eq:chemicalShift}, and the superfluid phase, where this symmetry is relaxed to the usual $\text U(1)_Q$ shifts Eq.~\eqref{eq:charge}. 
It is then natural to ponder whether intermediate fluid phases can exist that preserve only a subset of the infinite chemical shift symmetries, restricting the dynamics of the elementary charges on the comoving hypersurface to conserve only certain multipole moments.

In this section, we explore this possibility and propose a \emph{fracton fluid} phase, invariant under phase shifts linear in the comoving coordinates, corresponding to the comoving conservation of the dipole moment. An analogous \emph{fractonic solid} phase was recently proposed by Akash Jain \cite{Jain:2024ngx}\footnote{An idea to restrict the dynamics of charges with respect to the comoving observer was also discussed in \cite{Liang:2025gpp}.}. 
We begin by analysing the symmetry structure of fracton fluids and show that comoving multipole symmetries are generically incompatible with the full set of APDs, $\text{SDiff}(\mathbb{R}^2)$, which must therefore be restricted to its affine subgroup, $\text{SL}(2,\mathbb{R}) \ltimes \mathbb{R}^2$. Interestingly, the resulting symmetry group matches that of Ref.~\cite{PhysRevD.89.045002}, which implemented a nonlinear realization of this group to approximate an ideal fluid.

Next, we formulate a symmetry-invariant action that is manifestly covariant under general coordinate transformations, local $\text U(1)_Q$ shifts, as well as local comoving dipole shifts. From this action, we derive the associated nonlinear hydrodynamic equations, construct a Goldstone EFT and compute the dispersion relations of the low-energy excitations. In addition to the ordinary sound mode, we identify a propagating “second sound” mode with a magnonlike dispersion, $\omega \sim k^2$. Hence, the physical content of the theory matches with the fractonic fluids described thus far in the literature \cite{Grosvenor:2021rrt,Glorioso:2021bif,Glodkowski:2022xje,Armas:2023ouk,Jain:2023nbf,Glodkowski:2024ova,Glorioso:2023chm}, which conserve the dipole moment in physical space rather than in comoving space. In turn, the symmetry group implemented in \cite{PhysRevD.89.045002} is, at best, a crude approximation to ideal fluids, as it predicts additional gapless excitations that do not exist in a perfect fluid.

\subsection{Comoving dipole symmetry}
As discussed above, we define the fracton fluid phase by requiring invariance under linear shifts in the comoving space 
\be 
\Phi \rightarrow e^{i (\Lambda_0 +\Lambda_I \phi^I)} \Phi\,.
\ee 
In other words, we truncate the infinite-dimensional symmetry group of chemical shifts, $\text{CShift}(\infty)$, and consider its subgroup $\text{CShift}(1)$, consisting of constant and linear shifts.
Interestingly, by doing so, one must also restrict the $\text{SDiff}(\mathbb R^2)$ group down to the subgroup of affine area-preserving transformations $\text{SL}(2,\mathbb R) \ltimes \mathbb R^2$. 
To see this, consider the symmetry variations $\delta_{\Lambda_0}$ and $\delta_{\Lambda_1}$ associated to the comoving conservation of monopole and dipole moments. Then, using Eq.~\eqref{eq:algebraVariations} we have the following commutation relation 
\be \label{eq:fractonAlgebraSymm}
[\delta_{\Sigma_N}, \delta_{\Lambda_1}] = \delta_{\Lambda_{N-1}}\,, 
\ee 
with
\be 
\Lambda_{N-1} = \epsilon^{IJ} \Lambda_I \Sigma_{J J_2 \dots J_N} \phi^{J_2} \cdots \phi^{N_2}\,.  
\ee 
Therefore, to close the algebra we require $N \leq 2$ so that the $\text{SDiff}(\mathbb R^2)$ symmetry must be truncated to $\text{SL}(2,\mathbb R) \ltimes \mathbb R^2$. Note also that conservation of the $N$-th comoving multipole moment automatically implies conservation of all lower moments. 
The symmetry variations then furnish a Lie algebra 
\bs \label{eq:multipoleAlgebraVariations}
[\delta_{\Sigma_1}, \delta_{\Lambda_1}]  &= \delta_{\Lambda^\prime_0}  \,,  \quad [\delta_{\Sigma_2}, \delta_{\Lambda_1}]  = \delta_{\Lambda^\prime_1} \,, \\
[\delta_{\Sigma_1}, \delta_{\Sigma_2}]  &= \delta_{\Sigma^\prime_1}    \,, \quad [\delta_{\Sigma_2}, \delta_{ \Sigma_2^\prime}]  = \delta_{\Sigma^{\prime \prime}_2}  \,, 
\es 
with 
\bs
\Lambda^\prime_0 &= \epsilon^{IJ} \Lambda_I \Sigma_J\,, \\
\Lambda^\prime_1 &= \epsilon^{IJ} \Lambda_I \Sigma_{JK} \phi^K\,, \\
\Sigma^\prime_1 &= \epsilon^{IJ}  \Sigma_{IK}  \Sigma_J \phi^K \,, \\
\Sigma^{\prime \prime}_2 &=  \epsilon^{IJ}  \Sigma^\prime_{IK} \Sigma_{JL}  \phi^K \phi^L\,.
\es 
In the basis of generators Eq.~\eqref{eq:generatorsP}, the symmetry variations are 
\be 
\delta_{\Lambda_0} = i\Lambda_0 P^0\,, \quad  \delta_{\Lambda_1} = i\Lambda_I P^I\,, \quad \delta_{\Sigma_1} = \Sigma_I F^I\,, \quad \delta_{\Sigma_2} = \Sigma_{IJ} F^{IJ} \,.
\ee 
Then, the Lie algebra is specified with the following commutation relations
\be \begin{split}\label{eq:comovingDipoleAlgebra}
[F^I, P^J] &= -\epsilon^{IJ} P^0\,, \\
[F^{IJ}, P^K] &= -\frac{1}{2} \left( \epsilon^{IK} P^J + \epsilon^{JK} P^I \right)\,, \\
[F^{IJ}, F^K] &= -\frac{1}{2} \left( \epsilon^{IK} F^J + \epsilon^{JK} F^I \right)\,, \\
[F^{IJ}, F^{KL}] &= -\frac{1}{2}\left(
\epsilon^{IK} F^{JL}
+ \epsilon^{JK} F^{IL}
+ \epsilon^{IL} F^{JK}
+ \epsilon^{JL} F^{IK}
\right)\,.
\end{split}
\ee 
In Appendix \ref{app:matrixRepresentation}, we construct a four-dimensional matrix representation of the comoving dipole symmetry group
\be \label{eq:comovingDipoleSymmetry}
\left( \text{SL}(2,\mathbb R) \ltimes \mathbb R^2 \right) \ltimes \text{CShift}(1)\,,  
\ee 
acting on the internal vector space with coordinates $\phi^I$ and $\psi$, supplemented with an auxiliary dimension. 


\subsection{Relativistic model for fracton fluids}
We now implement the discussed symmetries and write down a field-theoretic model capturing the hydrodynamics of fracton fluids. In addition to the internal symmetries discussed above, we also impose Poincaré invariance so that the total symmetry group is 
\be \label{eq:fractonSymmetryGroup}
G = \text{ISO}(2,1) \times \Big( \left( \text{SL}(2,\mathbb R) \ltimes \mathbb R^2 \right) \ltimes \text{CShift}(1) \Big) \,.
\ee 
Moreover, we couple the theory to the spacetime metric $g_{\mu \nu}$, the monopole gauge field $A_\mu$, and dipole gauge field $A_{\mu I}$, rendering it covariant under diffeomorphisms of spacetime coordinates and local monopole and dipole gauge transformations, the latter acting as
\bs \label{eq:gaugeTransformations}
\Phi &\rightarrow e^{i(\Lambda_0(x) + \Lambda_I(x) \phi^I )} \Phi \,, \\
A_\mu &\rightarrow A_\mu + \partial_\mu \Lambda_0(x)\,, \\
A_{\mu I} &\rightarrow A_{\mu I} + \partial_\mu \Lambda_I(x)\,,
\es 
where the gauge parameters $\Lambda_0(x)$ and $\Lambda_I(x)$ are understood as local functions of spacetime.
Notice that the gauge fields $A_\mu$ and $A_{\mu I}$ transform under $\text{SL}(2,\mathbb R) \ltimes \mathbb R^2$. Indeed, the structure of the algebra Eq.~\eqref{eq:multipoleAlgebraVariations} implies
\be 
\delta_{\Sigma_1} A_\mu = \epsilon^{IJ} \Sigma_I A_{\mu J}\,, \quad \delta_{\Sigma_2} A_{\mu I} =  \Sigma_{IJ} \epsilon^{JK} A_{\mu K}\,.
\ee 
The variation of the dipole gauge field reflects the fact that $A_{\mu I}$ transforms in the fundamental representation of $\text{SL}(2,\mathbb R)$ and one can thus easily construct invariants by contracting comoving indices. However, the transformation of the monopole gauge field is nontrivial and requires some care. To circumvent this, we introduce an $\text{SL}(2,\mathbb R) \ltimes \mathbb R^2$--invariant combination
\be 
\mathcal A_\mu = A_\mu + \phi^I A_{\mu I}\,, 
\ee 
transforming as 
\be \label{eq:gaugeTransformationInvariantGaugeField}
\mathcal A_\mu \rightarrow \mathcal A_\mu + \partial_\mu \Lambda_0(x) + \phi^I \partial_\mu \Lambda_I(x)
\ee 
under local monopole and dipole gauge transformations. 

With these ingredients at hand, we are ready to formulate a field theory model for fracton fluids. 
Guided by \cite{Pretko:2018jbi,Jain:2024ngx} we propose the action
\be \label{eq:fractonAction}
S = \int d^3x \sqrt{-g} \, \mathcal L\,,
\ee 
with the Lagrangian
\be \label{eq:fractonLagrangian}
\mathcal L = \frac{i}{2}\big( \Phi^\dag D_0 \Phi - \Phi D_0 \Phi^\dag \big)  - \frac{\eta}{2} B^{IJ} B^{KL}  D_{IK}(\Phi, \Phi) D_{JL}(\Phi^\dag, \Phi^\dag)  - V(|\Phi|) + F(b) \,, 
\ee 
where we have introduced a covariant operator
\be \begin{split} \label{eq:covariantOperator}
D_{IJ}(\Phi, \Phi) &= \Phi D_{(I} D_{J)} \Phi  - D_I \Phi D_J \Phi - i   e^\mu_{(I} A_{|\mu| J)} \Phi^2 \,, \\
&=\Phi \Gamma^\mu_{IJ} D_\mu \Phi  + e^\mu_{(I} e^\nu_{J)} \big( \Phi D_\mu D_\nu \Phi  - D_\mu \Phi D_\nu \Phi \big) - i e^\mu_{(I} A_{|\mu| J)} \Phi^2 \,,
\end{split}
\ee 
with 
\be 
\Gamma^\mu_{IJ} \equiv e^\nu_{(I} \partial_\nu e^\mu_{J)}\,, \quad D_\mu \Phi \equiv \partial_\mu \Phi - i \mathcal A_\mu\,.
\ee 
Notice that $D_\mu \Phi$ is covariant under monopole gauge transformations, but under dipole gauge shifts it transforms nonlinearly
\be \label{eq:gaugeTransformationDerivative}
D_\mu \Phi \rightarrow e^{i \lambda_I(x) \phi^I } \left( D_\mu \Phi + i e^I_\mu \Lambda_I(x) \Phi \right)\,.
\ee 
Then, using Eqs.~\eqref{eq:gaugeTransformations}, \eqref{eq:gaugeTransformationInvariantGaugeField} and \eqref{eq:gaugeTransformationDerivative}, it is straightforward to verify that the derivative operator \eqref{eq:covariantOperator} is covariant under local dipole gauge transformations. 

We emphasize that the fracton fluid phase Eq.~\eqref{eq:fractonAction} is fundamentally different from the complex scalar field theories invariant under spacetime dipole symmetry \cite{Pretko:2018jbi,Seiberg:2019vrp,PhysRevResearch.2.023267,Bidussi:2021nmp,Jain:2021ibh,Jensen:2022ibn}.
In particular, the ordinary multipole symmetries, as classified by Gromov \cite{Gromov:2018nbv}, are incompatible with Lorentz and Galilean boosts whereas Eq.~\eqref{eq:fractonAction} is exactly invariant under Poincaré transformations\footnote{See also \cite{Bertolini:2022ijb,Afxonidis:2023pdq} for a Lorentz-covariant generalization of dipole symmetry based on the conservation of the four-dipole moment.}. Moreover, the theory Eq.~\eqref{eq:fractonAction} is covariantly coupled to the background Lorentzian geometry through the spacetime metric $g_{\mu \nu}$. On the other hand, dipole-conserving theories are notoriously difficult to reconcile with gravity and instead couple to Aristotelian background geometries \cite{Pena-Benitez:2021ipo,Pena-Benitez:2023aat,Jain:2021ibh,Grosvenor:2021hkn,Hartong:2024hvs,Afxonidis:2025wce}. 

The fracton fluid theory Eq.~\eqref{eq:fractonAction} is more closely related to the factonic solid phase introduced in Ref.~\cite{Jain:2024ngx} but differs in several key aspects. First, the fracton fluid is invariant under the $\text{SL}(2,\mathbb R) \ltimes \mathbb R^2$ symmetry group, in contrast to the $\text{SO}(2) \ltimes \mathbb R^2$ symmetry that characterizes an isotropic solid or ''jelly'' phase. This additional symmetry modifies the structure of the Lagrangian and, as a consequence, Eq.~\eqref{eq:fractonAction} does not support transverse phonon excitations. Second, the theory Eq.~\eqref{eq:fractonAction} is linear in time derivatives of $\Phi$ rather than quadratic. Therefore, the model \cite{Jain:2024ngx} exhibits a Higgs mode in the broken phase whereas the fracton fluid theory does not. Finally, in our formulation, the comoving multipole symmetries arise naturally in the description of the normal phase, and the fracton fluid is introduced as an interpolating state between the normal and superfluid phases. On the other hand, the corresponding ''crystal-multipole symmetries'' introduced in Ref.~\cite{Jain:2024ngx} are postulated from the outset.

We now proceed to recast the action Eq.~\eqref{eq:fractonAction} in a polar form $\Phi = \sqrt{\rho} e^{i \psi}$. 
First, let us observe that 
\bs 
D_\mu \Phi = \partial_\mu \sqrt{\rho} e^{i \psi} + i \mathcal D_\mu \psi \sqrt{\rho} e^{i \psi}\,,
\es 
where we have defined  
\be 
\mathcal D_\mu \psi = \partial_\mu \psi - \mathcal A_\mu\,,
\ee 
which transforms nonlinearly under dipole gauge transformations
\be 
\mathcal D_\mu \psi \rightarrow \mathcal D_\mu \psi + e^I_\mu \Lambda_I(x) \,.
\ee 
It is then possible to express the covariant derivative Eq.~\eqref{eq:covariantOperator} in terms of the polar variables 
\bs 
D_{IJ}(\Phi, \Phi) &= i e^{2 i \psi} \rho \Big[   \Gamma^\mu_{IJ}   \mathcal D_\mu \psi  +  e^\mu_{(I} e^\nu_{J)} \partial_\mu \mathcal D_\nu \psi    -  e^\mu_{(I} A_{|\mu| J)} \Big] \\ 
& +  e^{2 i \psi} \Big[ \Gamma^\mu_{IJ} \sqrt{\rho} \partial_\mu \sqrt{\rho} + e^\mu_I e^\nu_J \big(  \sqrt{\rho} \partial_\mu \partial_\nu \sqrt{\rho}  - \partial_\mu \sqrt{\rho}   \partial_\nu \sqrt{\rho}   \big) \Big]\,.
\es 
Since we are interested in the hydrodynamic regime, the ``quantum pressure'' contributions involving derivatives of $\rho$ can be neglected. This can be justified by solving for $\rho$ from its equations of motion and truncating the resulting series expansion, following steps analogous to the superfluid case (see Eq.~\eqref{eq:rhoSuperfluid}). Instead, we proceed more crudely here and simply drop all derivatives of $\rho$ outright.
Then, the Lagrangian reads
\be \label{eq:polarFractonAction}
S = \int d^3x \sqrt{-g}  \Big[ -\rho \mathcal D_0 \psi - \frac{\lambda}{2} \rho^2 - \frac{\eta}{2}  B^{IJ} B^{KL}  \rho^2 \mathcal D_{IK} \psi \mathcal D_{JL} \psi   + F(b) \Big]\,,
\ee 
where we have defined 
\bs 
\mathcal D_0 \psi &\equiv u^\mu \mathcal D_\mu \psi \,, \\
\mathcal D_{IJ} \psi &\equiv \Gamma^\mu_{IJ}   \mathcal D_\mu \psi  +  e^\mu_{(I} e^\nu_{J)} \partial_\mu \mathcal D_\nu \psi    -  e^\mu_{(I} A_{|\mu| J)} \,.
\es 
From the equation of motion for the auxiliary variable $\rho$ we obtain the relation 
\be \label{eq:rhoRelationFracton}
\mathcal D_0 \psi = - \rho \left( \lambda + \eta  B^{IJ} B^{KL}   \mathcal D_{IK} \psi \mathcal D_{JL} \psi \right) \,, 
\ee 
which can be used to express $\rho$ as a perturbative series in $\psi$. However, we find it more useful to keep the relation exact and use it to trade $\mathcal D_0 \psi$ for $\rho$ in what follows.

\subsection{Hydrodynamic equations}
In this section we derive the hydrodynamic equations for fracton fluids.

We begin by varying the fracton fluid action Eq.~\eqref{eq:polarFractonAction} with respect to the theory's fields and external sources 
\bs 
\delta S = \int d^3x \sqrt{-g} \Big[ - \frac{1}{2} T_{\mu \nu} \delta g^{\mu \nu} + J^\mu \delta \mathcal A_\mu + J^{\mu I} \delta A_{\mu I}  + \mathcal C_I \delta \phi^I + \mathcal K \delta \psi  \Big] \,.
\es  
Requiring invariance under infinitesimal diffeomorphisms and local gauge monopole and dipole transformations yields the following Ward identities
\bs 
    \nabla_\nu T^{\mu \nu} &= \mathcal F^{\mu \nu} J_\nu + F^{\mu \nu I} J_{\nu I} +  \mathcal C_I e^{I \nu} + \mathcal K  \partial^\nu \psi  \,, \\ 
\nabla_\mu J^\mu &=  \mathcal K   \,, \\ \nabla_\mu J^{\mu I} &= - e^I_\mu J^\mu +  \mathcal K \phi^I \,.
\es 
The full dynamical content of the theory is encoded in the equations of motion $\mathcal K = 0$ and $ \mathcal C_I = 0$. Equivalently, one may express the same information through the continuity equations
\bs \label{eq:equationsFracton}
    \nabla_\nu T^{\mu \nu} &= \mathcal F^{\mu \nu} J_\nu + F^{\mu \nu I} J_{\nu I}  \,, \\ 
\nabla_\mu J^\mu &=  0  \,, \\ \nabla_\mu J^{\mu I} &= - e^I_\mu J^\mu  \,.
\es 
Applying a number of variational formulas (see Appendix \ref{app:geometricVariations} for details) to the action Eq.~\eqref{eq:polarFractonAction} we obtain the constitutive relations for the currents 
\bs \label{eq:constitutiveFracton}
J^\mu &= \rho u^\mu + \eta \rho^2 B^{IJ} B^{KL} \Big[ \Gamma^\mu_{IK} \mathcal D_{JL} \psi - \nabla_\nu \Big( e^\mu_{(K} e^\nu_{L)}  \mathcal D_{JL} \psi \Big) \Big]\,, \\
J^{\mu I} &=  \eta \rho^2 B^{IJ} B^{KL} e^\mu_L \mathcal D_{JK} \psi \,, \\
T^{\mu \nu} &= 2 \eta      \rho^2 \xi^{\mu \rho} \xi^\nu_\rho  - \big( \rho \mathcal D_0 \psi  + F_b b\big)  u^\mu u^\nu + \big(\mathcal L-F_b b\big) g^{\mu \nu}\,,
\es 
where in the last line we have introduced a ''superfluid velocity''
\be 
\xi_{\mu \nu} = e^I_\mu e^J_\nu \mathcal D_{IJ} \psi\,,
\ee 
with the norm defined as $\xi^2 = D_{\mu \nu} \psi \mathcal D^{\mu \nu} \psi$. 

Matching to the standard form, $T^{\mu \nu} \simeq \cdots + \left(\epsilon + p \right) u^\mu u^\nu + p g^{\mu \nu}$, we can read off the thermodynamic pressure and energy density functions
\bs 
p &= - \frac{\rho}{2} \mathcal D_0 \psi  - \frac{\eta}{2}    \rho^2 \mathcal \xi^2 + F - F_b b \,, \\
\epsilon &= - \frac{\rho}{2} \mathcal D_0 \psi + \frac{\eta}{2}    \rho^2 \xi^2 - F\,.
\es 
Imposing the relation $\epsilon + p = \mu \rho + T s$ implies $\mu = - \mathcal D_0 \psi$, $T = -F_b$ and $s = b$.  

Finally, using the relation Eq.~\eqref{eq:rhoRelationFracton} we find that the pressure obeys the thermodynamic relation 
\be 
dp = n d\mu + s dT + \frac{\eta \rho^2}{4} d \xi^2 \,. 
\ee 

\subsection{Excitation spectrum}
In this section, we study the hydrodynamic modes of fracton fluids. We identify two propagating modes--one with the linear (soundlike) and second with quadratic (magnonlike) dispersion. 

Following the discussion presented in Sec.~\ref{sec:equilibrium}, we consider a homogenous equilibrium configuration Eq.~\eqref{eq:equilibrium} and place the system at a finite chemical potential\footnote{In principle, it is also possible to switch on a finite chemical potential for the comoving dipole symmetry $A_{\mu I} = (\mu_I, 0)$.}, $\mathcal A_\mu = (\mu_0, 0)$. Then, the most general stationary equilibrium state minimizing the effective potential Eq.~\eqref{eq:effectivePotential} is 
\be 
\Phi = \sqrt{\rho_0} e^{i(c_0 + c_I \phi^I)}\,,
\ee 
where $c_0$ and $c_I$ are arbitrary constants, which we can fix to zero for our convenience. Altogether, the equilibrium configuration is characterized by the following symmetry breaking pattern
\be \label{eq:fractonSSB}
\text{ISO}(2,1) \times \Big( \left( \text{SL}(2,\mathbb R) \ltimes \mathbb R^2 \right) \ltimes \text{CShift}(1) \Big) \rightarrow \left( \text{SO}(2) \ltimes \mathbb R^2 \right) \times \mathbb R \,.
\ee 

We know proceed to study the hydrodynamic modes of the system and determine the dispersion relations of the collective excitations. To this goal, we consider linearized fluctuations around the equilibrium state. Plugging the expansion Eq.~\eqref{eq:fluctations} into the Lagrangian Eq.~\eqref{eq:polarFractonAction} we arrive at the effective Lagrangian
\be \label{eq:effectiveFractons}
\mathcal L = \frac{1}{2 \lambda} (\partial_t \varphi)^2   + \rho_0 \partial_t \pi^i \partial_i \varphi - \frac{\eta }{2} \rho_0^2 (\partial_i \partial_j \varphi)^2 + \frac{w_0}{2} (\partial_t \pi^i)^2 +\frac{F_{bb} s_0^2}{2} (\partial_i \pi^i)^2 \,.
\ee
In writing Eq.~\eqref{eq:effectiveFractons} we have made use of the identities collected in Appendix~\ref{app:linearized}, dropped terms that are higher than quadratic in the Goldstone fields, integrated out $\delta \rho$ and defined equilibrium enthalpy density $w_0 = \rho_0 \mu_0 - F_b s_0 $. 

Comparing the fracton EFT with the superfluid theory Eq.~\eqref{eq:superfluidEffective}, we observe that the Goldstone $\varphi$ enters with two spatial derivatives, which is a consequence of the comoving dipole symmetry.

To identify the theory's modes we perform a Fourier transformation of the effective action, yielding 
\be
S =\frac{1}{2} \int \frac{ d^2 k d\omega }{(2\pi)^{3/2}}   \begin{pmatrix}
    \tilde \varphi(k,\omega) \\
    \tilde \pi_{||}(k,\omega) \\
    \tilde \pi_{\perp}(k,\omega) 
\end{pmatrix}^T \begin{pmatrix}
    \frac{1}{\lambda} \omega^2 - \eta \rho_0^2  k^4 & \rho_0 \omega k & 0\\
    \rho_0 \omega k & w_0 \omega^2 + F_{bb} s_0^2 k^2 & 0\\
    0  & 0 & w_0 \omega^2
\end{pmatrix}\begin{pmatrix}
   \tilde \varphi(-k,-\omega) \\
   \tilde \pi_{||}(-k,-\omega) \\
   \tilde \pi_{\perp}(-k,-\omega) 
\end{pmatrix}\,.
\ee 
The characteristic equation admits the solutions 
\bs \label{eq:fractonModes}
\omega_s &= \pm c_s k\,, \quad c_s^2 = \frac{\mu_0 \rho_0 - s_0^2 F_{bb}}{w_0}\,, \\
\omega_m &= \pm v_m k^2\,, \quad v_m^2 = \eta  s_0 \rho_0 \frac{-F_{bb}}{\mu_0 \rho_0 - s_0^2 F_{bb}}\,.
\es 
The sound mode $\omega_s$ corresponds to a collective excitation carrying both the charge $\psi$ and thermal component $\pi^i$ and corresponds to the normal sound mode in ordinary charged fluids Eq.\eqref{eq:sound}. On the other hand, the magnonic mode $\omega_m$ only involves an oscillation in the phase field $\psi$. Finally, the shear mode is still non-propagating $\omega_\perp = 0$ due to the $\text{SL}(\mathbb R, 2)$ symmetry, which forbids the appearance of a kinetic term for the shear component. Therefore, relaxing the APD symmetry $\text{SDiff}(\mathbb R^2)$ down to the subgroup of area-preserving affine transformations $\text{SL}(2,\mathbb R) \ltimes \mathbb R^2$ still describes a fluid phase rather than a solid.

The spectrum of low-energy excitations is in qualitative agreement with previous studies on s-wave fracton superfluids with dipole moment conserved in physical space \cite{Armas:2023ouk,Glodkowski:2024ova,Jain:2023nbf,Jain:2024kri}. However, contrary to previous approaches utilizing the hydrodynamic paradigm, our derivation follows directly from the microscopic model Eq.~\eqref{eq:fractonAction}. The fractonic two-fluid model Eq.~\eqref{eq:fractonAction}, incorporating both the fractonic charge sector and the normal (thermal) component, generalizes the zero-temperature fractonic superfluid phases \cite{PhysRevResearch.2.023267} to a finite temperature regime.

The authors of Ref.~\cite{PhysRevD.89.045002} employed the same symmetry group Eq.~\eqref{eq:fractonSymmetryGroup} and symmetry-breaking pattern Eq.~\eqref{eq:fractonSSB} to model an ideal fluid as a nonlinear realization thereof. Our analysis of the microscopic model with the same symmetries shows, however, that this symmetry group predicts an additional gapless excitation with quadratic dispersion, which has no analogue in ordinary fluids. This indicates that the fractonic symmetry group Eq.~\eqref{eq:fractonSymmetryGroup} does not describe ordinary charge-carrying fluids, which must instead exhibit invariance under the full chemical shift symmetry, as discussed around Eq.~\eqref{eq:chemicalShift}.

\section{Conclusion}\label{sec:conclusion}
In this work, we have presented a theory of charged fluids via a comoving hypersurface approach, with the complex matter field defined thereon. The proposed models, Eqs.~\eqref{eq:normalAction} and \eqref{eq:superfluidTheory}, serve as a UV completion to the EFTs of charged fluids \cite{Dubovsky:2011sj} and finite-temperature superfluids \cite{Nicolis:2011cs}, respectively. 

We have demonstrated from first principles that charge-carrying fluids exhibit the restrictive chemical shift symmetry, Eq.~\eqref{eq:chemicalShift}, which renders the elementary charges immobile on the comoving plane, thereby realizing fractonic phenomenology in a concrete physical system. This fractonic symmetry forbids a kinetic term for the phase field, trivializing its dispersion and leaving only a single longitudinal sound mode in the low-energy spectrum, as befits ordinary fluids. The absence of chemical shift symmetry in the superfluid phase leads to the appearance of a second sound mode.

Our framework provides a natural interpolation between normal fluids, whose charges are completely immobile on the comoving plane, and superfluids, whose charges are fully mobile. Such intermediate fluid phases are characterized by charges with partially restricted mobility on the comoving hypersurface, respecting a finite subgroup of chemical shifts corresponding to the comoving conservation of certain multipole moments. Focusing on the simplest case of fluids with comoving dipole symmetry, we have proposed a fracton fluid phase, in analogy with the fractonic solid phase introduced in Ref.~\cite{Jain:2024ngx}. We have shown that the physical content of the fracton fluid theory Eq.~\eqref{eq:fractonAction} accurately reflects the low-energy spectrum of fracton fluids studied extensively in the literature \cite{Armas:2023ouk,Glodkowski:2024ova,Jain:2023nbf,Jain:2024kri}, which respect dipole conservation in physical space. 

We conclude by outlining some interesting open problems for future exploration. 
Perhaps the most pressing question is the generalization of the comoving framework to include dissipative effects. Formulating dissipative hydrodynamics within an action principle allows for a systematic account of stochastic fluctuations, enables the computation of higher-point correlation functions, and ensures consistency with fluctuation–dissipation relations. While notable progress has been achieved \cite{Kovtun:2014hpa,Crossley:2015evo,Haehl:2018lcu,Jain:2023obu,Mullins:2025vqa,Firat:2025upx}, existing formulations are technically demanding and, in practice, difficult to use beyond reproducing established results, except in simple cases such as diffusive systems \cite{Delacretaz:2023ypv,Delacretaz:2023pxm}. In addition, such formulations typically rely on additional gauge symmetries whose physical provenance is not transparent. It would therefore be interesting to construct a dissipative EFT in terms of a complex scalar defined on the Schwinger–Keldysh contour over two copies of the comoving hypersurface. An open question is whether such a formulation reproduces all first-order dissipative transport coefficients, or whether some are missing due to the restricted field content of the comoving formulation.

Another interesting direction is the potential adaptation of our construction to spinful fluids, which are of particular relevance for applications in heavy-ion collisions. Such systems are most naturally formulated in the tetrad formalism, coupled to the spin connection, possibly within a torsionful geometry \cite{Gallegos:2021bzp,Gallegos:2022jow}. To incorporate spin, one could introduce a set of fermionic fields, defined on the comoving hypersurface, in analogy with the bosonic matter fields discussed in this work. Notably, spinful fluids exhibit massive modes \cite{Hongo:2021ona}, which may possibly be traced back to a more microscopic description within the comoving formalism.

Finally, it would be interesting to modify our framework to incompressible quantum hall fluids and obtain a field-theoretical description of such phases, along the lines of \cite{Son:2013rqa,Geracie:2014nka}. For this purpose, one would need to replace the Poincaré symmetry used here with Galilean symmetry and place a system in a homogeneous background magnetic field. Subsequently, one could consider a formal lowest Landau level limit by sending the mass to zero $m \rightarrow 0$, which should imply the incompressibility constraint.

\begin{acknowledgments}
I am indebted to Piotr Surówka, Paweł Matus, Francisco Peña-Benítez, Sergej Moroz and Eren Firat for useful discussions and comments. A.G. is supported in part by the Polish National Science Centre (NCN) Sonata Bis grant 2019/34/E/ST3/00405.
\end{acknowledgments}

\appendix

\section{Variational formulae}\label{app:variational}
In this Appendix, we collect some variational formulae, which are needed to derive the constitutive relations presented in the main part. We also provide the expressions for linearized fluctuations around a stationary fluid configuration. 

\subsection{Geometric variations}\label{app:geometricVariations}
Let us begin by collecting the formulas required to derive Eq.~\eqref{eq:stressTensor}. For this purpose, we vary $b$ with respect to the metric $g_{\mu \nu}$ and comoving fields $\phi^I$,
\be \label{eq:variationb}
\delta b = \frac{b}{2} B_{IJ} \delta B^{IJ} =  \frac{b}{2} \Pi_{\mu \nu} \delta g^{\mu \nu} + b e^\mu_I \delta e^I_\mu\,, 
\ee 
where we have used 
\be \label{eq:varyMetric}
\delta B^{IJ} = e^I_\mu e^J_\nu \delta g^{\mu \nu} + 2 e^{J \mu} \delta e^I_\mu\,.
\ee 
We will also need the standard formula for the variation of the metric determinant 
\be
\delta \sqrt{-g} =  -\frac{1}{2} \sqrt{-g} g_{\mu \nu} \delta g^{\mu \nu}\,. 
\ee 
To derive the constitutive relations for charged fluids Eq.~\eqref{eq:constitutiveNormal} we need to vary $D_0 \psi = u^\mu (\partial_\mu \psi - A_\mu)$ with respect to the metric $g_{\mu\nu}$ and gauge field $A_\mu$. First, let us compute the variation of the velocity field, $u^\mu = \frac{v^\mu}{b}$, where $v^\mu$ is defined in Eq.\eqref{eq:vector} and its variation reads
\be 
\delta v^\mu = \frac{1}{2} v^\mu g_{\nu \rho} \delta g^{\nu \rho} \,.
\ee 
Then, using \eqref{eq:variationb} we obtain 
\bs 
\delta u^\mu &= -\frac{1}{2} u^\mu u_\nu u_\rho \delta g^{\nu \rho}\,.
\es 
It is then straightforward to verify 
\be
\delta D_0 \psi = -\frac{1}{2}  D_0 \psi u_\mu u_\nu \delta g^{\mu \nu} - u^\mu \delta A_\mu \,. 
\ee
%
%
To evaluate the superfluid currents Eqs.~\eqref{eq:conservationSuperfluid} and \eqref{eq:stressTensorSuperfluid} we also used $\delta D_\mu \psi = -  \delta A_\mu$ and 
\be 
\delta \Pi^{\alpha \beta} = \left(  \delta^\alpha_\rho \delta^\beta_\sigma - u^\alpha u^\beta u_\rho u_\sigma  \right) \delta g^{\rho \sigma}\,,
\ee 
which follows straightforwardly
from the definition \eqref{eq:projector}. 

Finally, to compute the fracton currents Eqs.~\eqref{eq:constitutiveFracton} we need Eq.~\eqref{eq:varyMetric} and also the variations 
\bs 
\delta \mathcal D_0 \psi &= -\frac{1}{2} \mathcal D_0 \psi u_\mu u_\nu \delta g^{\mu \nu} - u^\mu \delta \mathcal A_\mu  \,, \\
\delta \mathcal D_{IJ} \psi &= -\Gamma^\mu_{IJ}    \delta \mathcal A_\mu -  e^\mu_{(I} e^\nu_{J)} \partial_\mu \delta \mathcal A_\nu    -  e^\mu_{(I} \delta A_{|\mu| J)} \,.
\es

\subsection{Linearized fluctuations}\label{app:linearized}
In deriving the EFTs Eqs.~\eqref{eq:normalFluidEFT}, \eqref{eq:superfluidEffective}, and \eqref{eq:effectiveFractons} in the main text, we performed an expansion around a homogeneous background 
\be 
\phi^I = \sqrt{s_0} \delta^I_i \big( x^i + \pi^i \big)\,,
\ee 
and truncated the resulting expansion at quadratic order in fluctuations. In this appendix, we collect the expansions that were used in the derivation.

%

In order to derive Eq.~\eqref{eq:normalFluidEFT} we need to expand $b$ and $u^\mu$. Assuming Minkowski metric, we find  
\be \begin{split} \label{eq:linearizedExp}
    b &= s_0 \Big( 1 + \partial_i \pi^i +\frac{1}{2} \epsilon^{ij} \epsilon_{ab} \partial_i \pi^a \partial_j \pi^b -\frac{1}{2}(\partial_t \pi^i)^2 + \dots \Big)\,, \\
u^0 &= 1 + \frac{1}{2}(\partial_t \pi^i)^2 +\dots\,, \\
u^i &= -\partial_t \pi^i +\partial_t \pi^i \partial_j \pi^j +\epsilon^{ij} \epsilon_{ab} \partial_j \pi^a \partial_t \pi^b+\dots
\end{split}
\ee
Using the expression for velocity we also have 
\be \label{eq:otherLinearized}
D_0 \varphi = \partial_t \varphi - \mu_0 - \frac{\mu_0}{2} (\partial_t \pi^i)^2 - \partial_t \pi^i \partial_i \varphi + \dots 
\ee 
Then, the effective theory Eq.~\eqref{eq:normalFluidEFT} follows straightforwardly after substituting \eqref{eq:linearizedExp} and \eqref{eq:otherLinearized} into the Lagrangian Eq.~\eqref{eq:normalActionPolar}.

For the superfluid EFT, Eq.~\eqref{eq:superfluidEffective}, we also need the expansion of the projector Eq.~\eqref{eq:projector}, given by 
\be \begin{split}
\Pi^{00} &= \frac{1}{4}(\partial_t \pi^i)^4 + \dots  \\
\Pi^{0i} &= -\partial_t \pi^i + \dots  \\ 
\Pi^{ij} &= \delta^{ij} + \partial_t \pi^i \partial_t \pi^j + \dots
\end{split}
\ee 
%
%
Finally, to derive the EFT for fracton fluids Eq.~\eqref{eq:effectiveFractons} we make use of the following expansions 
\bs 
\mathcal D_0 \psi &\simeq \partial_t \psi - \mu_0 - \frac{\mu_0}{2} (\partial_t \pi^i)^2 - \partial_t \pi^i \partial_i \psi \,, \\
\mathcal D_{IJ} \psi & \simeq  \delta^\mu_I \delta^\nu_J \partial_\mu \partial_\nu \psi \,, \\
B^{IJ} &\simeq \delta^I_i \delta^J_j \delta^{ij}\,.
\es 

\section{Representation of the comoving dipole symmetry group}\label{app:matrixRepresentation}
In this appendix we realize the comoving dipole symmetry group Eq.~\eqref{eq:comovingDipoleSymmetry} as a linear action on an extended field space by embedding $\left(\phi^1, \phi^2, \psi \right)$ into $\mathbb R^4$. In particular, we work on the space with coordinates 
$( \phi^1, \phi^2, \psi, 1)$ so that the three-dimensional field space is modelled as a hypersurface in $\mathbb R^4$ with the last auxiliary coordinate fixed to unity.
In this setup, a generic element of the comoving dipole symmetry group $g \in G$ is represented by the matrix $\rho(g) \in \text{GL}(4,\mathbb R)$ of the form   
\be 
\rho(g) = \begin{pmatrix}
   \text M & 0 & \mathbf b \\
    \mathbf \Lambda & 1 & \Lambda_0 \\
        0 & 0 & 1 
\end{pmatrix}  \,, \quad \mathbf b = \begin{pmatrix} b_1 \\ b_2 \end{pmatrix}, \quad
\mathbf \Lambda = \begin{pmatrix} \Lambda_1 & \Lambda_2 \end{pmatrix}\,,
\ee 
where $M \in \mathrm{SL}(2,\mathbb R)$ can be expressed as a product of rotation, squeeze, and (horizontal) shear transformations
\be 
M(\theta, a, \alpha) = R(\theta) \cdot S(a)  \cdot H(\alpha)\,,
\ee 
with
\be \label{eq:SLMatrices}
R(\theta) = \begin{pmatrix}
    \cos{\theta} & -\sin{\theta} \\
    \sin{\theta} & \cos{\theta} 
\end{pmatrix} \,, \quad S(a) = \begin{pmatrix}
    e^a & 0 \\
    0 & e^{-a}
\end{pmatrix} \,, \quad H(\alpha) = \begin{pmatrix}
    1 & \alpha \\
    0 & 1 
\end{pmatrix}\,.
\ee 
The action of the comoving dipole group is realized linearly on the coordinates 
\be 
\Psi \rightarrow \rho(g) \Psi\,, \quad \Psi \equiv 
( \phi^1, \phi^2, \psi, 1)^T\,.
\ee 
The set of matrices $\rho(g)$ forms a subgroup of $\text{GL}(4,\mathbb R)$, and is a Lie group in its own right. An element of the corresponding Lie algebra can be expressed as 
\be 
\mathfrak{g} = \begin{pmatrix}
   \text m & 0 & \mathbf b \\
    \mathbf \Lambda & 0 & \Lambda_0 \\
        0 & 0 & 0 
\end{pmatrix} \,,
\ee 
where $\text m  =  \text r(\theta ) +   \text s(a) +  \text h(\alpha) \in \mathfrak{sl}(2,\mathbb R)$
with 
\be 
\text r(\theta ) = \theta \begin{pmatrix} 0 & -1 \\ 1 & 0 \end{pmatrix}\,, \quad
\text s(a) = a \begin{pmatrix} 1 & 0 \\ 0 & -1 \end{pmatrix}\,, \quad
\text h(\alpha) = \alpha \begin{pmatrix} 0 & 1 \\ 0 & 0 \end{pmatrix}\,,
\ee 
forming a basis of $\mathfrak{sl}(2,\mathbb R)$ consisting of generators associated to matrices Eq.~\eqref{eq:SLMatrices}.
In terms of the full Lie algebra basis, we have
\begin{equation}
\mathfrak{g} = \theta X_r + a X_s + \alpha   X_h + b^I P_I + \Lambda^I Q_I + \Lambda_0 Q\,,
\end{equation}
where $X_r, X_s, X_h$ generate the $\mathfrak{sl}(2,\mathbb{R})$ sector, $P_I$ represent translations in the comoving space, $Q_I$ are comoving dipole generators, and $Q$ is the central charge generating constant phase shifts.
It is then straightforward to verify the following commutation relations 
\bs \label{eq:algebraComovingAppendix}
[P_I, Q_J] &= - \delta_{IJ} Q_0\,, &\quad  [X_s,X_h] &= 2  X_h\,, \\
[X_r,X_s] &= 2 X_r + 4 X_h\,, &\quad  [X_r,X_h] &= - X_s\\
[X_r, P_I] &= \epsilon_{IJ} P_J\,, &\quad [X_r, Q_I] &= \epsilon_{IJ} Q_J\,,  \\
  [X_s, P_1] &= P_1 \,, &\quad [X_s, P_2] &= - P_2\,, \\
 [X_s, Q_1] &= -Q_1 \,, &\quad [X_s, Q_2] &=  Q_2\,, \\
[X_h, P_2] &= P_1 \,, &\quad  [X_h, Q_1] &= - Q_2 \,.
\es 
This is precisely the algebra presented in the main text Eq.~\eqref{eq:comovingDipoleAlgebra} albeit written in a different basis. Notice that \eqref{eq:algebraComovingAppendix} contains as a subalgebra a dipole algebra \cite{Gromov:2018nbv} 
\be 
[P_I, Q_J] = - \delta_{IJ} Q_0\,, \quad 
[X_r, P_I] = \epsilon_{IJ} P_J\,, \quad [X_r, Q_I] = \epsilon_{IJ} Q_J\,.
\ee 
The first commutator encodes the fact that dipole moment is generically charged under translations (and vice versa) whereas the remaining two reflect the fact that $P_I$ and $Q_I$ transform as vectors under rotations generated by $X_r$.

For fracton fluids, the rotational symmetry $\text{SO}(2)$ is enlarged to $\text{SL}(2,\mathbb R)$, whose generators satisfy, in our chosen basis, the commutation relations
\be 
[X_s,X_h] = 2  X_h\,, \quad
[X_r,X_s] = 2 X_r + 4 X_h\,, \quad  [X_r,X_h] = - X_s\,.
\ee 
Finally, the remaning commutation relatioms 
\bs 
  [X_s, P_1] &= P_1 \,, &\quad [X_s, P_2] &= - P_2\,, \\
 [X_s, Q_1] &= -Q_1 \,, &\quad [X_s, Q_2] &=  Q_2\,, \\
[X_h, P_2] &= P_1 \,, &\quad  [X_h, Q_1] &= - Q_2 \,,
\es 
specify how momentum and dipole generators transform under squeezing and horizontal shearing. 

 \bibliographystyle{JHEP}
 \bibliography{biblio.bib}

\end{document}